\documentclass[a4paper,11pt]{article}

\usepackage[totalwidth=410pt, totalheight=590pt]{geometry}
\usepackage{cite}
\usepackage{amsmath}
\numberwithin{equation}{section}

\newcommand {\beq} {\begin{equation}}
\newcommand {\eeq} {\end{equation}}
\newcommand {\beqa} {\begin{eqnarray}}
\newcommand {\eeqa} {\end{eqnarray}}
\newcommand {\beqan} {\begin{eqnarray*}}
\newcommand {\eeqan} {\end{eqnarray*}}

\newcommand {\nn} {\nonumber}
\newcommand {\ph}[1]{\phantom{#1}}
\newcommand {\ie}{i.e.~}
\newcommand {\eg}{e.g.~}
\newcommand {\cf}{cf.~}

\newcommand {\sss} {\scriptscriptstyle}

\newcommand{\calF}{\ensuremath{\mathcal{F}}}

\newcommand{\al}{\ensuremath{\alpha}}
\newcommand{\be}{\ensuremath{\beta}}
\newcommand{\ga}{\ensuremath{\gamma}}

\newcommand{\de}{\ensuremath{\delta}}
\newcommand{\De}{\ensuremath{\Delta}}
\newcommand{\eps}{\ensuremath{\epsilon}}
\newcommand{\veps}{\ensuremath{\varepsilon}}
\newcommand{\ka}{\ensuremath{\kappa}}
\newcommand{\la}{\ensuremath{\lambda}}

\newcommand{\La}{\ensuremath{\Lambda}}
\newcommand{\si}{\ensuremath{\sigma}}
\newcommand{\Si}{\ensuremath{\Sigma}}
\newcommand{\om}{\ensuremath{\omega}}
\newcommand{\Om}{\ensuremath{\Omega}}
\newcommand{\Th}{\ensuremath{\Theta}}

\newcommand{\alh}{\ensuremath{\hat{\alpha}}}
\newcommand{\beh}{\ensuremath{\hat{\beta}}}
\newcommand{\gah}{\ensuremath{\hat{\gamma}}}
\newcommand{\deh}{\ensuremath{\hat{\delta}}}
\newcommand{\muh}{\ensuremath{\hat{\mu}}}
\newcommand{\nuh}{\ensuremath{\hat{\nu}}}
\newcommand{\rhoh}{\ensuremath{\hat{\rho}}}
\newcommand{\sih}{\ensuremath{\hat{\si}}}

\newcommand{\kde}[2]{\de_{#1}^{\ph{#1}#2}}
\newcommand{\dia}[3]{{#1}_{#2}^{\ph{#2}#3}}

\newcommand{\desi}{\ensuremath{\delta_{\sss \Si}}}

\newcommand{\mathHb}[1]{{\mathop{\kern0pt#1}\limits^{\,\sss
      \prime\prime}\vphantom{#1}}}

\newcommand{\com}[2]{\left[ #1 , #2 \right]}
\newcommand{\acom}[2]{\left\{ #1 , #2 \right\}}

\newcommand {\we} {\wedge}
\newcommand {\pa} {\partial}

\newcommand{\eqnlab}[1]{\label{eqn:#1}}

\newcommand{\eqnref}[1]{(\ref{eqn:#1})}

\newcommand{\Eqnref}[1]{Eq.~(\ref{eqn:#1})}

\newcommand{\Eqsref}[1]{Eqs.~(\ref{eqn:#1})}

\newcommand{\seclab}[1]{\label{sec:#1}}

\newcommand{\Secref}[1]{Section~\ref{sec:#1}}

\author{P\"ar Arvidsson}

\begin{document}

 \pagestyle{empty}

\begin{center}

\vspace*{2cm}

\noindent
{\LARGE\textsf{\textbf{Superconformal symmetry in the interacting
      theory
      \\[5mm] of $(2,0)$ tensor multiplets and self-dual strings}}}
\vskip 3truecm

{\large \textsf{\textbf{P\"ar Arvidsson}}} \\
\vskip 1truecm
{\it Department of Fundamental Physics\\ Chalmers University of
  Technology \\ SE-412 96 G\"{o}teborg,
  Sweden}\\[3mm] {\tt par.arvidsson@chalmers.se} \\
\end{center}
\vskip 1cm
\noindent{\bf Abstract:}
We investigate the concept of superconformal symmetry in six
dimensions, applied to the interacting theory of $(2,0)$ tensor
multiplets and self-dual strings. The action of a superconformal transformation on the superspace coordinates is found, both from a six-dimensional perspective and by using a superspace with eight bosonic and four fermionic dimensions.
The transformation laws for all
fields in the theory are derived, as well as general expressions for the
transformation of on-shell superfields. Superconformal invariance is
shown for the interaction of a self-dual string with a background
consisting of on-shell tensor multiplet fields, and we also find an
interesting relationship between the requirements of superconformal
invariance and those of a local fermionic $\kappa$-symmetry. Finally, we try to construct a superspace analogue of the Poincar\'e dual to the string world-sheet and consider its properties under superconformal transformations.

\newpage
\pagestyle{plain}

\tableofcontents

\section{Introduction}

One of the most interesting discoveries in string/$M$-theory during
the past decade is without doubt the six-dimensional $(2,0)$
theories, named after the supersymmetry algebra under which they are
invariant~\cite{Nahm:1978}. These are superconformal quantum theories without dynamical gravity, and first appeared in a compactification of Type~IIB string
theory on a four-dimensional hyper-K\"ahler
manifold~\cite{Witten:1995}. They obey an $ADE$ classification
(see~\cite{Henningson:2004} for an intrinsically six-dimensional
motivation for why the simply laced Lie algebra series $A$, $D$ and
$E$ appear) but have no other discrete or continuous
parameters. However, the theories have a moduli space parametrized by
the expectation values of a set of scalar fields.

There is a second origin for the $A$-series of these theories in terms
of $M$-theory, where they arise as the world-volume theory on a stack
of parallel $M5$-branes~\cite{Witten:1996,Dasgupta:1996}. More
specifically, a stack of $r+1$ branes yields the $A_r$ version of
$(2,0)$ theory, where $r$ denotes the rank of the associated Lie
algebra. The fluctuations of the $M5$-branes are described by $r$ so
called $(2,0)$ tensor multiplets and the $5r$ moduli correspond to the
transverse distances between the branes. It is also possible for
membranes to stretch between two
$M5$-branes~\cite{Strominger:1996,Townsend:1996,Becker:1996}; the
intersections will then appear as self-dual strings from the
six-dimensional world-volume perspective on the $M5$-brane. We get in
total $r(r+1)/2$ different species of such strings, corresponding to
the number of ways to connect the branes. The string
tension is proportional to the distance between the branes in
question, and is therefore related to the moduli of the
theory. Specifically, if the branes coincide, the strings become
tensionless. This picture yields in a simple way the different degrees
of freedom of $(2,0)$ theory; for a more complete discussion, see
e.g.~\cite{AFH:2002}.

An intrinsically six-dimensional formulation of the $(2,0)$ theories is
still lacking, and it is the ultimate goal of our research to
find such a definition. In our previous work,
we have pursued a program where we consider the theory at a point away
from the origin of its moduli space, where the strings are
tensile. This introduces a scale in the theory and breaks the
conformal invariance spontaneously, but also provides a basis for
doing perturbation theory, since in the limit of large string tension,
the theory describes the well understood free $(2,0)$ tensor multiplet
and free self-dual strings. The dimensionless expansion parameter will
then be the square of some typical energy divided by the string tension. For
simplicity, we have chosen to work in the $A_1$ version of
$(2,0)$ theory, which includes a single type of string and a single tensor
multiplet.

In this paper, we consider the hitherto uninvestigated superconformal
symmetry of our model. It is well known that superconformal field theories cannot exist in space-times with more than six dimensions. Moreover, in six dimensions, the largest possible supersymmetry consistent with superconformal invariance is $\mathcal{N}=(2,0)$. This means that $(2,0)$ theory, regarded as a superconformal theory, has the largest possible supersymmetry in the highest possible dimension. This observation alone provides a motivation for studying these theories.

The aim of the present paper is firstly to provide a framework for the study of superconformal invariance in six-dimensional $(2,0)$ theory by using superfields. Secondly, we want to investigate whether our model~\cite{AFH:2003coupled} for a self-dual string interacting with a tensor multiplet background is superconformally invariant, and in what sense. Finally, we would like to use the new superconformal tools to find some clues on how to formulate the full interacting theory, \ie when the tensor multiplet fields do not obey their free equations of motion.

It is our intention to make the paper self-contained, thereby reviewing and summarizing some well-known results on conformal and superconformal symmetry. We think that it is worthwhile to include these, in order to understand how the superfield transformations work and how they are derived. It is also, in some cases, non-trivial to adapt the known results to our notations and conventions.


The outline of this paper is as follows: \Secref{bosonic} is devoted to the study of (bosonic) conformal transformations in six dimensions, as a warm-up exercise preceding the superconformal model. We find explicit expressions for the action of the conformal group on the space-time coordinates, both from a six-dimensional perspective and by considering a projective hypercone in eight dimensions, where the conformal transformations act linearly. We also study how different fields behave under conformal transformations, especially the ones in the bosonic sector of the $(2,0)$ tensor multiplet. The results of \Secref{bosonic} are not new, but still useful from a pedagogical point of view.

\Secref{susy} discusses superconformal symmetry. As in the previous section, we first find how the transformations act on coordinates, this time in a superspace with six bosonic and sixteen fermionic dimensions. The coordinate transformations are also found by considering a projective supercone in a higher-dimensional superspace; a method which, to the best of our knowledge, has not appeared previously in the literature. Next, we use a superfield formalism as a way of compactifying and simplifying the notation and derive how superconformal transformations act on the superfields of the complete $(2,0)$ tensor multiplet and on self-dual spinning strings. The explicit transformation laws are used to show that the model describing a string interacting with a tensor multiplet background is superconformally invariant. We find the new and non-trivial result that the requirements of superconformal symmetry and those of a local fermionic $\kappa$-symmetry impose similar restrictions on the possible coupling terms, thereby indicating the uniqueness of the theory.

Finally, \Secref{poincare} discusses the construction of a superspace analogue of the Poincar\'e dual to the string world-sheet and its properties under superconformal transformations.

\section{Conformal invariance in the bosonic theory}
\seclab{bosonic}

Before discussing the issue of superconformal invariance, it is
worthwhile to consider a simpler case, which still contains some
important aspects of the problem. Therefore, we begin by working with
the model describing a spinnless self-dual string interacting with
the bosonic part of the $(2,0)$ tensor multiplet in six
dimensions~\cite{AFH:2002}. The results in this section have appeared previously in the literature, but it is nevertheless useful to review some aspects in order to get a complete picture. We will also refer (for comparison) to formulae in this section later in this paper, to point out similarities and differences between the bosonic and the superconformal models.

\subsection{Conformal coordinate transformations}
\seclab{conf_bos}

In this subsection, we consider the action of conformal transformations on the coordinates $x^\mu$, $\mu=0,\ldots,5$, in a six-dimensional space-time. These transformations are called \emph{passive}, meaning that they act on the coordinates themselves, rather than on the fields of the theory.

Conformal transformations act on space-time in such a way that the \emph{angle} between two intersecting curves is left invariant. This means that the infinitesimal proper time interval
\beq
d \tau^2 = - \eta_{\mu \nu} dx^\mu dx^\nu
\eeq
transforms according to
\beq
d \tau^2 \rightarrow \Om^2(x) d \tau^2,
\eeq
where $\eta_{\mu \nu}$ is the flat space-time metric (with "mostly plus" signature) and $\Om(x)$ denotes a space-time dependent quantity that also involves the parameters of the conformal transformation. For an infinitesimal transformation, such that $\Om(x)=1+\La(x)$, this yields that
\beq
\eqnlab{proptimetransf}
\de (d \tau^2) = 2 \La(x) d \tau^2,
\eeq
where $\La(x)$ is an infinitesimal function.

It is well known that the most general infinitesimal coordinate transformation that respects \Eqnref{proptimetransf} is given by
\beq
\eqnlab{x_transf_bosonic}
\de x^{\mu} = a^{\mu} + \om^{\mu \nu} x_{\nu} +
\la x^{\mu} + c^{\mu} x^2 - 2 c\cdot x x^{\mu},
\eeq
where $c\cdot x \equiv \eta_{\mu \nu} c^{\mu} x^{\nu} = c_\nu
x^\nu$. The (constant) parameters $a^\mu$, $\om^{\mu\nu}$, $\la$
and $c^\mu$ are related to the different transformations of the
conformal group according to the following table: \vspace{3mm}
  \begin{center}
    \begin{tabular}{l|l|l}
      {\bf symmetry} & {\bf generator} & {\bf parameter} \\ \hline
      translations & $P_{\mu} = \pa_{\mu}$ & $a^{\mu}$ \\
      rotations and Lorentz boosts & $M_{\mu \nu} = x_{[\nu}
      \pa_{\mu]}$ & $\om^{\mu \nu}$ \\
      dilatations & $D=x^{\mu} \pa_{\mu}$ & $\lambda$ \\
      special conformal transformations & $K_{\mu}= x^2 \pa_{\mu} - 2
      x_{\mu} x^{\nu} \pa_{\nu} $ & $c^{\mu}$ \\
    \end{tabular}
  \end{center}
\vspace{3mm} where the differential expressions for the generators make the equation
\beq
\de x^{\mu} = \left( a^{\nu} P_{\nu} + \om^{\nu \rho} M_{\nu
  \rho} + \la D + c^{\nu} K_{\nu} \right) x^{\mu}
\eqnlab{delta_x_bos}
\eeq
valid. In total, we have 28 parameters, in agreement with the dimensionality of the conformal group in six dimensions, denoted by $SO(6,2)$.

From \Eqnref{x_transf_bosonic}, it is easily shown that the differential $dx^\mu$ transforms according to
\beq
\eqnlab{transf_dx}
\de (dx^\mu) = \left(\om^{\mu\nu} + 4 c^{[\mu} x^{\nu]} \right) dx_\nu + \left(\la - 2 c \cdot x \right) dx^\mu,
\eeq
which implies that the proper time interval indeed transforms according to \Eqnref{proptimetransf} with $\La(x)=\la - 2 c \cdot x$.

Leaving the coordinate transformations aside for a while, we turn to the commutation relations relating the generators of the conformal group to eachother. By direct calculation, it is easily verified that the differential operators defined in the table above obey
\beq
\eqnlab{bosonic_comm_coord}
\begin{array}{rclcrcl}
\com{P_{\mu}}{M_{\nu \rho}} & = & - \eta_{\mu [\nu} P_{\rho]} & \qquad &
\com{P_{\mu}}{D} & = & P_{\mu} \\
\com{K_{\mu}}{M_{\nu \rho}} & = & - \eta_{\mu [\nu} K_{\rho]} & \qquad
& \com{K_{\mu}}{D} & = & -K_{\mu} \\
\com{P_{\mu}}{K_{\nu}} & = & - 4 M_{\mu \nu} - 2 \eta_{\mu \nu} D & \quad &
\com{M_{\mu \nu}}{D} & = & 0 \\
\com{M_{\mu \nu}}{M^{\rho \si}} & = & 2 \de_{[\mu}^{\ph{[\mu} [\rho}
      M_{\nu]}^{\ph{\nu]} \si]}. & && \\
\end{array}
\eeq
This defines the conformal group $SO(6,2)$ in six dimensions. The
$SO(6,2)$ structure may be made more explicit by introducing a new set of
generators $J_{\muh \nuh}$, $\muh,\nuh = 0,\ldots,7$, according to
\beq
  \begin{array}{rcl}
  M_{\mu\nu} & = & J_{\mu \nu} \\
  P_\mu & = & 2 \left( J_{7 \mu} + J_{6 \mu} \right) \\
  K_{\mu} & = & 2 \left( J_{7 \mu} - J_{6 \mu} \right) \\
  D & = & 2 J_{76}.
  \end{array}
  \eqnlab{J_bos}
\eeq
These obey the Lorentz-type commutation relations
\beq
\eqnlab{J_munu}
\com{J_{\muh\nuh}}{J_{\rhoh\sih}} = \eta_{\rhoh[\muh}
  J_{\nuh]\sih} - \eta_{\sih[\muh} J_{\nuh]\rhoh},
\eeq
where the metric $\eta_{\muh \nuh} =
\mathrm{diag}(-1,1,1,1,1,1,1,-1)$. Thus, the conformal transformations
act as rotations in an eight-dimensional space with two time-like
directions.

There is a formulation (first suggested by Dirac~\cite{Dirac:1936}, whose
work was later continued by Kastrup~\cite{Kastrup:1966} and by
Mack and Salam~\cite{Mack:1969}) in which a
$d$-dimensional space-time is regarded as a projective hypercone in a
space with one extra space-like and one extra time-like dimension, just like
the one suggested in the previous paragraph.
In the higher-dimensional space, conformal symmetry acts linearly, \ie
as a rotation with the generator $J_{\muh\nuh}$. Denote the coordinates in
this space by $y^{\muh}$, and define the projective hypercone by
\beq
\eqnlab{bos_hypercone}
y^2 = \eta_{\muh \nuh} y^{\muh} y^{\nuh} = 0.
\eeq
This equation is clearly Lorentz invariant in the eight-dimensional space,
and therefore conformally invariant from a six-dimensional point of view.

Next, let the infinitesimal rotation parameters be assembled in the
matrix $\pi^{\muh \nuh}$, such that
\beq
\de y^{\muh} = \pi^{\rhoh\sih} J_{\rhoh \sih} y^{\muh} =
  \pi^{\muh\sih} y_{\sih}.
\eqnlab{delta_y}
\eeq
The last equality follows from the obvious definition
\beq
J_{\muh\nuh} = y_{[\nuh} \pa_{\muh]},
\eeq
where the derivative acts on the $y$-variables.

To make contact with the six-dimensional space-time, we impose that
\beq
\pi^{\rhoh\sih} J_{\rhoh \sih} = \om^{\rho \si} M_{\rho \si} + a^\rho
  P_\rho + c^\rho K_\rho + \la D,
\eeq
which when combined with \Eqnref{J_bos} yields that
\beq
  \begin{array}{rcl}
  \om^{\mu\nu} & = & \pi^{\mu \nu} \\
  a^\mu & = & \frac{1}{2} \left( \pi^{7 \mu} + \pi^{6 \mu} \right) \\
  c^{\mu} & = & \frac{1}{2} \left( \pi^{7 \mu} - \pi^{6 \mu} \right) \\
  \la & = & \pi^{76}.
  \end{array}
  \eqnlab{pi_bos}
\eeq
Next, we parametrize the projective hypercone by
\beqa
y^{\mu} & = & \ga x^\mu \nn \\
\eqnlab{hypercone_param}
y^7-y^6 & = & \ga \\
y^7+y^6 & = & \ga \eta_{\rho\si} x^\rho x^\si, \nn
\eeqa
where $\ga$ is related to the projectiveness of the hypercone and therefore
$\ga \neq 0$.

Under a conformal transformation as in \Eqnref{delta_y}, it turns out that the
quantity $x^\mu$ introduced in \Eqnref{hypercone_param} must transform exactly as the coordinates $x^\mu$ in \Eqnref{x_transf_bosonic}, while $\ga$ transforms according to
\beq
\de \ga = \left( 2 \eta_{\mu\nu} c^\mu x^\nu - \la \right) \ga.
\eeq
This shows that the surface of the projective hypercone defined in
\Eqnref{bos_hypercone} behaves as the six-dimensional space-time we started
with. In this way, we have derived the conformal coordinate transformation in \Eqnref{x_transf_bosonic} from a higher-dimensional perspective, a viewpoint that will be useful later in this paper as well. The eight-dimensional formulation also suggests a way of formulating manifestly conformally invariant quantities, an observation that hopefully will be pursued in future work.

\subsection{Conformal field transformations}

Having discussed the conformal group and its action on the six-dimensional space-time thoroughly, we now turn to the question of how the conformal group generators act on fields defined over this space-time. This action consists of two parts: one due to the dependence of the field in question on the space-time coordinate
$x^\mu$ and one due to specific properties of the field itself. The latter part
consists of the action of the stability subgroup of $x=0$ (the little group) on the fields, \eg the transformation properties of a vector field under
Lorentz transformations. We will (in contrast to our convention in the previous subsection) adopt the \emph{active} view on conformal
transformations, meaning that the space-time coordinates are fixed but
the fields change upon such a transformation. By considering the
stability subgroup mentioned before, it can be found~\cite{Mack:1969}
that a general field $\phi^i(x)$ (where $i$ denotes some index or
indices) transforms according to
\beqa
\de_{\sss C} \phi^i(x) & = & a^{\mu} \pa_{\mu} \phi^i(x) + \om^{\mu
  \nu} \left[ x_{\nu} \pa_{\mu} \phi^i(x) + \left( \Si_{\mu \nu} \phi
\right)^i(x) \right] + {} \nn \\
& & {} + \la \Big[ x^{\mu} \pa_{\mu} \phi^i(x) + w
\phi^i(x) \Big] + c^{\mu} \Big[ x^2 \pa_{\mu} \phi^i(x) - 2 x_{\mu} x^{\nu}
\pa_{\nu} \phi^i(x) + {} \nn \\
\eqnlab{mack_transf}
& & {} + 4 x^{\nu} \left( \Si_{\mu \nu} \phi \right)^i(x)
- 2 x_{\mu} w \phi^i(x) + (k_{\mu} \phi)^i(x) \Big],
\eeqa
where $\Si_{\mu\nu}$, $w$ and $k_\mu$ are defined through
\beq
  \begin{array}{rcl}
  \com{M_{\mu\nu}}{\phi^i(0)} & = & \left( \Si_{\mu \nu} \phi
  \right)^i(0) \\[2mm]
  \com{D}{\phi^i(0)} & = & w \phi^i(0) \\[2mm]
  \com{K_{\mu}}{\phi^i(0)} & = & (k_{\mu} \phi)^i(0).
  \end{array}
\eeq
Here, $M_{\mu\nu}$, $D$ and $K_\mu$ should not be confused with the
differential expressions in the table above; in these equations (when
acting on fields), they denote the full generators of Lorentz
transformations, dilatations and special conformal transformations,
respectively.

Thus, $\Si_{\mu\nu}$ and $k_\mu$ correspond to the intrinsic
properties of the field under space-time rotations and special
conformal transformations, respectively, while $w$ is
the conformal weight of the field. Together, these generate the
stability subgroup of $x=0$. We also note that a \emph{primary} field
corresponds to one having $k_\mu=0$.

It is convenient to summarize the transformation~\eqnref{mack_transf} in the
expression
\beq
\eqnlab{general_transf}
\de_{\sss C} \phi^i(x) = \xi^{\mu}(x) \pa_{\mu} \phi^i(x) + \Om^{\mu
  \nu}(x) \left( \Si_{\mu \nu} \phi \right)^i(x) + \La(x) w
\phi^i(x) + c^{\mu} \left(k_{\mu} \phi\right)^i(x),
\eeq
where the space-time dependent parameter functions are defined by
\beqa
\eqnlab{xi_bos}
\xi^{\mu}(x) & \equiv & a^{\mu} + \om^{\mu \nu} x_{\nu} + \la x^{\mu} +
c^{\mu} x^2 - 2 c \cdot x x^{\mu} \\
\eqnlab{Om_bos}
\Om^{\mu \nu}(x) & \equiv & \om^{\mu \nu} + 4 c^{[\mu} x^{\nu]} \\
\eqnlab{La_bos}
\La(x) & \equiv & \la - 2 c \cdot x.
\eeqa
This expression is also stated in~\cite{Claus:1998} and will prove to be
useful when dealing with superfields and the superconformal group later in this paper. Note that
the expression for $\xi^\mu(x)$ in \Eqnref{xi_bos} coincides with the
expression for $\de x^\mu$ in \Eqnref{x_transf_bosonic},
indicating that this part corresponds to the change in the field due
to its dependence on the space-time coordinates.

When acting on \emph{fields}, the generators obey the same commutation relations as in~\Eqnref{bosonic_comm_coord}, but with the opposite sign; this is due to the difference between active and passive transformations.

To conclude, the conformal
transformation of a general field is specified by stating its
conformal weight ($w$), its behaviour under Lorentz transformations
($\Si_{\mu\nu}$) and its properties under special conformal transformations
($k_\mu$).

\subsection{The bosonic tensor multiplet}

The next step is to apply the results of the previous section to
the fields in the bosonic sector of the $(2,0)$ tensor multiplet, consisting of five scalar fields and a two-form gauge potential. We denote the scalar
fields, transforming in the vector representation of the $SO(5)$
$R$-symmetry group, by the antisymmetric matrix $\phi^{ab}$,
$a,b=1,\ldots,4$, satisfying the algebraic condition
\beq
\Om_{ab} \phi^{ab} = 0,
\eeq
where $\Om_{ab}$ is the $SO(5)$ invariant antisymmetric
tensor. Obviously, $a$ and $b$ are $SO(5)$ spinor indices.
The two-form potential is denoted $b$ and has an associated field
strength $h=db$, both $R$-symmetry scalars. Actually, it is only the
self-dual part of the field strength that is part of the tensor
multiplet, but in order to give a Lagrangian description of the
theory~\cite{Witten:1997}, we include the anti self-dual part as a
spectator field. It is essential to keep this part decoupled when
adding interactions to the theory. The field content, along with the
reality properties, is more thoroughly discussed
in~\cite{AFH:2003free}.

The fields have the following properties under conformal
transformations:
\beq
\begin{array}{rclcrclcrcl}
(\Si_{\mu \nu} \phi^{ab}) & = & 0 & \quad & (k_{\mu} \phi^{ab}) & = & 0 & \quad
  & w_{\sss \phi} & = & 2 \\
(\Si_{\mu \nu} b)_{\rho \si} & = & \eta_{\mu[\rho} b_{\si]\nu} -
\eta_{\nu[\rho} b_{\si]\mu} & & (k_{\mu} b)_{\rho \si} & = & 0 & &
  w_{\sss b} & =  & 2.
\end{array}
\eeq
The conformal weights $w_{\sss \phi}$ and $w_{\sss b}$ are bounded
from below by a unitarity condition~\cite{Minwalla:1998,Dobrev:2002},
and due to the BPS property of the tensor multiplet representation,
this bound is saturated. The weights coincide with the mass dimensions
of the fields.

Using the expression \eqnref{general_transf}, this means that the
fields of the bosonic theory transform according to
\beqa
\eqnlab{transf_phi}
\de_{\sss C} \phi^{ab}(x) & = & \xi^{\mu}(x) \pa_{\mu} \phi^{ab}(x) + 2 \La(x)
\phi^{ab}(x) \\
\eqnlab{transf_b}
\de_{\sss C} b_{\mu \nu}(x) & = & \xi^{\rho}(x) \pa_{\rho} b_{\mu
  \nu}(x) + 2 \Om_{[\mu}^{\ph{[\mu} \rho}(x) b_{\nu] \rho}(x) + 2 \La(x)
  b_{\mu \nu}(x) \\
\eqnlab{transf_h}
\de_{\sss C} h_{\mu \nu \rho}(x) & = & \xi^{\si}(x) \pa_{\si} h_{\mu
  \nu \rho}(x) - 3 \Om_{[\mu}^{\ph{\mu} \si}(x) h_{\nu \rho] \si}(x) + 3
\La(x) h_{\mu \nu \rho}(x),
\eeqa
where we have required the conformal transformation operator to
commute with the exterior derivative, in the sense that
\beq
\eqnlab{bos_der_comm}
d (\de_{\sss C} b) = \de_{\sss C} (db) = \de_{\sss C} h.
\eeq
Note that this conformal transformation operator is defined in an
abstract sense, acting in different ways on different
fields. Specifically, its explicit form is different when acting on the
derivative of a field than on the field itself. The
relation~\eqnref{bos_der_comm} may seem evident here, but it is useful
to relate to it when we discuss the superconformal transformations
later in this paper.

Comparing \Eqsref{transf_b} and \eqnref{transf_h}, we see that both $b$ and its exterior derivative $h=db$ are primary fields, meaning that they have no $k_{\mu}$ piece in their transformation laws, \cf \Eqnref{general_transf}. It should be noted that, in general, derivatives of primary fields are not necessarily primary, consider \eg the variation
\beqa
\de_{\sss C} \left( \pa_{\mu} \phi^{ab}(x) \right) & = &  \pa_{\mu}
\left( \de_{\sss C} \phi^{ab}(x) \right) \\
& = & \xi^{\nu}(x) \pa_{\nu} \pa_{\mu}
\phi^{ab}(x) - \Om_{\mu}^{\ph{\mu} \nu}(x) \pa_{\nu} \phi^{ab}(x) + {}
\nn \\
\eqnlab{transf_dphi}
& & {} + 3 \La(x)
\pa_{\mu} \phi^{ab}(x) - 4 c_{\mu} \phi^{ab}(x),
\eeqa
where we see that the generators of the conformal group act as expected on a field with a subscript vector index and mass dimension 3, apart from the non-primary piece with parameter $c_{\mu}$. Thus, although $\phi^{ab}$ is a primary field, its derivative is not.

The bosonic model also includes self-dual strings, described by an
embedding field $X^\mu$, which is a function of the world-sheet
coordinates (labelled $\si^i$, $i=1,2$) and transforms as a Lorentz
vector. In particular, since we have adopted the active viewpoint on
conformal transformations, it is natural to conclude that $X^\mu$
transforms according to
\beq
\eqnlab{transf_X}
\de_{\sss C} X^{\mu} = - \xi^{\mu}(X) = - a^{\mu} -
\om^{\mu}_{\ph{\mu} \nu} X^{\nu} - \la X^{\mu} - c^{\mu} X^2 + 2
c \cdot X X^{\mu},
\eeq
\ie in the same way as the space-time coordinate $x^\mu$ in
\Eqnref{x_transf_bosonic}, but with the opposite sign. This expression
does not look as nice as the transformations of the tensor multiplet
fields found above, but if we instead consider the variation of the differential
$\tilde{d} X^{\mu} \equiv d\si^i \pa_i X^\mu$ (which is what we will need in
explicit calculations), we get that
\beq
\eqnlab{transf_dX}
\de_{\sss C} (\tilde{d}X^{\mu}) =  - \Om^{\mu}_{\ph{\mu} \nu}(X) \tilde{d} X^{\nu} -
\La(X) \tilde{d} X^{\mu},
\eeq
in accordance with the expected transformation of a field with a
vector index, having conformal weight $w_{\sss \tilde{d}X}=-1$. Note also, from
\Eqsref{general_transf} and \eqnref{transf_X}, that all terms apart
from those involving $\Si_{\mu\nu}$, $w$ and $k_\mu$ vanish manifestly
when transforming the pull-back of a space-time field to the
world-sheet of the string.

Using the transformation rules found above, it is a straight-forward task to prove that the interaction described in~\cite{AFH:2002} is conformally invariant. It should be
stressed that this model, because of its electromagnetic coupling,
suffers from a classical anomaly~\cite{Henningson:2004}. This anomaly
is cancelled when fermionic degrees of freedom are included in the
model in a supersymmetric way.

\section{Conformal invariance in the supersymmetric theory}
\seclab{susy}

In this section we turn to the complete theory, incorporating fermionic
degrees of freedom in the model. We take advantage of the superfield notation in
order to simplify expressions and keep them similar and comparable to the bosonic results in the previous section. This section mixes results that have appeared elsewhere in the literature with new findings. It should also be mentioned that, in some cases, our derivation methods are quite different from those that can be found in previous papers on this subject.

\subsection{Including spinors in the model}
\seclab{ferm_conv}

In this subsection, we introduce the notations and conventions used for
spinors and state the \emph{bosonic} conformal transformations of the
\emph{fermionic} tensor multiplet fields. It summarizes old results, but is included for completeness to keep the paper self-contained.

The $(2,0)$ tensor multiplet includes, apart from the scalars and the
two-form potential mentioned in the previous section, four chiral
spinors transforming in the spinor representation of
the $SO(5)$ $R$-symmetry group. These are denoted as $\psi^a_{\al}$
where $\al=1,\ldots,4$ is an $SO(5,1)$ Weyl spinor index. An anti-Weyl
spinor is denoted by a superscript $\al$ index. The spinor fields obey
a symplectic Majorana reality condition~\cite{Kugo:1983}.

It is convenient to use spinor indices also for the bosonic
representations of the Lorentz group. This means that we let (by contracting with the appropriate gamma matrices)
\beqa
x^\mu & \rightarrow & x^{\al \be} = x^{[\al \be]} \\
\pa_\mu & \rightarrow & \pa_{\al \be} = \pa_{[\al \be]} \\
\eqnlab{metric_spinor}
\eta_{\mu \nu} & \rightarrow & \frac{1}{2} \eps_{\al\be\ga\de} \\
\eqnlab{b->b}
b_{\mu \nu} & \rightarrow & \dia{b}{\al}{\be} \textrm{, such that }
\dia{b}{\al}{\al}=0 \\
h_{\mu \nu \rho} + (*h)_{\mu \nu \rho}  & \rightarrow & h_{\al \be} =
h_{(\al \be)} \\
h_{\mu \nu \rho} - (*h)_{\mu \nu \rho}  & \rightarrow & h^{\al \be} =
h^{(\al \be)}.
\eeqa
Note that the last two equations conveniently separate the self-dual
and anti self-dual parts of the three-form $h$ into different
representations of the Lorentz group.

As is indicated by \Eqnref{metric_spinor}, pairs of antisymmetric
indices may be raised and lowered according to
\beq
\pa^{\al \be} = \frac{1}{2} \eps^{\al \be \ga \de} \pa_{\ga \de}.
\eeq
Finally, we introduce the dot product between vectors in the same way
as before, such that
\beq
\pa \cdot x \equiv \pa_{\al \be} x^{\al \be} = 6.
\eeq

After these preliminaries, let us return to the conformal group and
its generators.
Naively, the bosonic differential generators of the table in
\Secref{conf_bos} are translated into \vspace{3mm}
  \begin{center}
    \begin{tabular}{l|l}
      {\bf generator} & {\bf parameter} \\ \hline
      $P_{\al \be} = \pa_{\al \be}$ & $a^{\al \be}$ \\
      $M_{\al \be;\ga \de} = \frac{1}{2} \left( x_{\ga \de} \pa_{\al
      \be}- x_{\al \be} \pa_{\ga \de} \right) $ & $\om^{\al \be;\ga
      \de}$ \\
      $D=x^{\al \be} \pa_{\al \be}$ & $\lambda$ \\
      $K^{\al \be}= x^2 \pa^{\al \be} - 2
      x^{\al \be} x^{\ga \de} \pa_{\ga \de} $ & $c_{\al \be}$, \\
    \end{tabular}
  \end{center}
\vspace{3mm} but it is more convenient to introduce a dual notation
for the Lorentz generator $M$ and its corresponding parameter $\om$,
according to
\beqa
\dia{M}{\al}{\be} & = & \eps^{\be\ga\de\veps}M_{\al\ga;\de\veps}
\\
\dia{\om}{\al}{\be} & = & \frac{1}{2} \eps_{\al\ga\de\veps}
\om^{\be\ga;\de\veps},
\eeqa
in analogy with the notation for the two-form $b$ in \Eqnref{b->b}.
This yields that
\beq
\om_{\be}^{\ph{\be}\al} M_{\al}^{\ph{\al}\be} = \om^{\al\be;\ga\de}
  M_{\al\be;\ga\de},
\eeq
and the differential operator becomes
\beq
M_{\al}^{\ph{\al}\be} = x^{\be\ga} \pa_{\al \ga} - x_{\al \ga}
\pa^{\be \ga} = 2 x^{\be\ga} \pa_{\al \ga} - \frac{1}{2}
\kde{\al}{\be} x \cdot \pa,
\eeq
which is traceless as required.

Under a conformal transformation, the spinor field $\psi^a_\al$ transforms
according to
\beq
\eqnlab{transf_psi}
\de_{\sss C} \psi^a_{\al} = \xi^{\ga\de}(x) \pa_{\ga\de}
\psi^a_{\al} + \dia{\Om}{\al}{\ga}(x) \psi^a_{\ga} + \frac{5}{2}
\La(x) \psi^a_{\al},
\eeq
where the
$x$-dependent parameter functions are the obvious translations from
\Secref{conf_bos} and become
\beqa
\xi^{\al \be}(x) & = & a^{\al \be} + \dia{\om}{\ga}{\al} x^{\ga \be} +
\dia{\om}{\ga}{\be} x^{\al \ga} + (\la - 2 c \cdot x ) x^{\al \be} +
c^{\al \be} x^2 \\
\eqnlab{Om_bos_al}
\dia{\Om}{\al}{\be}(x) & = & \dia{\om}{\al}{\be} - 4 c_{\al \ga} x^{\be
  \ga} + c \cdot x \kde{\al}{\be}  \\
\eqnlab{La_bos_al}
\La(x) & = & \la - 2 c \cdot x,
\eeqa
in terms of which the transformations \eqnref{transf_phi} and
\eqnref{transf_h} are
\beqa
\eqnlab{transf_phi_al}
\de_{\sss C} \phi^{ab} & = & \xi^{\ga\de}(x) \pa_{\ga\de}
\phi^{ab} + 2 \La(x) \phi^{ab} \\
\eqnlab{transf_h_al}
\de_{\sss C} h_{\al \be} & = & \xi^{\ga \de}(x) \pa_{\ga \de} h_{\al
  \be} + \dia{\Om}{\al}{\ga}(x) h_{\ga\be} +
  \dia{\Om}{\be}{\ga}(x) h_{\al\ga} + 3 \La(x) h_{\al\be}.
\eeqa

This ends the discussion concerning the conformal group with only bosonic
generators. In the next subsection, we extend the model to incorporate
fermionic generators, thereby forming the full superconformal group.

\subsection{Superconformal coordinate transformations}

Since we want to treat a supersymmetric model, it is convenient to
supplement the six bosonic space-time coordinates $x^{\al \be}$ by a
set of sixteen fermionic coordinates $\theta^{\al}_a$,
see~\cite{AFH:2003coupled} for a more thorough introduction of these
concepts. The fermionic coordinates are anticommuting, \ie Grassmann
odd, as usual and transform as anti-Weyl spinors under Lorentz rotations and as spinors under $R$-symmetry transformations.

Following the logic of the previous section, the obvious next step is to find out what the generators of the conformal group $SO(6,2)$ look like in this \emph{superspace} and how they act on the coordinates. Evidently, the generators must change in comparison to the bosonic case, in order to incorporate the non-trivial action of the conformal group on the fermionic coordinates $\theta^{\al}_a$. For example, since $\theta^{\al}_a$ transforms as an anti-Weyl spinor under Lorentz rotations, the corresponding generator ($\dia{M}{\al}{\be}$) must be modified. The commutation relations in \Eqnref{bosonic_comm_coord} are expected to remain unchanged as expressed in terms of abstract generators.

However, we want to go a bit
further and also include the $R$-symmetry group $SO(5)$, generated by
$U^{ab}=U^{(ab)}$, and supersymmetry, which is generated by the fermionic
$Q^a_{\al}$. We are then forced to introduce the generators
$S_a^{\al}$ of \emph{special supersymmetry} as well, which arise as the commutator of
the supersymmetry and the special conformal symmetry generators. Altogether, we have then
arrived at the superconformal group $OSp(8^*|4)$, which is the
expected symmetry group of the complete theory. We will denote the
parameters of supersymmetry, $R$-symmetry and special supersymmetry
transformations by $\eta^\al_a$, $v_{ab}$ and $\rho^a_\al$,
respectively. Note that $Q^a_{\al}$, $S_a^{\al}$, $\eta^\al_a$ and
$\rho^a_\al$ all are fermionic (Grassmann odd) quantities.

In this subsection, we will content ourselves with the action of the
superconformal group on the coordinates in superspace, postponing
their action on fields to the next subsection. Starting from
the purely bosonic parts of the generators of $SO(6,2)$ ($P_{\al\be}$, $\dia{M}{\al}{\be}$, $D$
and $K^{\al\be}$) obtained above and the well-known
generator of supersymmetry ($Q^a_\al$), we can make suitable ans\"atze for
the unknown but essential additional parts (including fermionic variables
and derivatives) of these
generators as well as for the newly introduced generators ($S^\al_a$ and
$U^{ab}$). By requiring these generators to obey certain of the
commutation relations of the $OSp(8^*|4)$ group (which is related to the requirement that the algebra should close), the unknown
coefficients may be determined and the resulting differential
generators are found to be
 \beqa
  \eqnlab{super_P}
  P_{\al \be} & = & \pa_{\al \be} \\
\dia{M}{\al}{\be} & = & 2 x^{\be\ga} \pa_{\al \ga} - \frac{1}{2} \kde{\al}{\be} x \cdot \pa + \theta^{\be}_c \pa^c_\al - \frac{1}{4} \dia{\de}{\al}{\be} \theta \cdot \pa \\
  D & = & x \cdot \pa + \frac{1}{2} \theta \cdot \pa \\
  K^{\al \be} & = & -4 x^{\al\ga} x^{\be\de} \pa_{\ga\de}
  - \Om^{ab} \Om^{cd} \theta_a^{\ga}
  \theta_b^{[\al} \theta_c^{\be]} \theta_d^{\de} \pa_{\ga\de} + {}
  \nn \\
  & & {}+2 \theta_c^{[\al} \Big( 2 x^{\be] \ga} - i \Om^{ab}
  \theta_a^{\be]} \theta_b^{\ga} \Big) \pa^c_{\ga} \\
  \eqnlab{Q_super}
  Q^a_{\al} & = & \pa^a_{\al} - i \Om^{ac} \theta_c^{\ga} \pa_{\al\ga}
  \\
  S_a^{\al} & = & \Om_{ac} \Big( 2 x^{\al \ga} - i \Om^{bd}
  \theta_b^{\al} \theta_d^{\ga} \Big) \pa^c_{\ga} + 2i
  \theta_a^{\ga} \theta_c^{\al} \pa^c_{\ga} - {} \nn \\
  & & {} - i \theta_a^{\ga} \Big( 2 x^{\de \al} - i
  \Om^{bd} \theta_b^{\de} \theta_d^{\al} \Big) \pa_{\ga \de} \\
  \eqnlab{super_U}
  U^{ab} & = & \frac{1}{2} \Big( \Om^{ac} \theta_c^{\ga}
  \pa_{\ga}^b + \Om^{bc} \theta_c^{\ga} \pa_{\ga}^a \Big),
 \eeqa
where the dot product between two fermionic quantities $a^\al_a$ and $b^b_\be$ is defined by
\beq
a \cdot b \equiv a^\ga_c b^c_\ga.
\eeq
Note that the fermions involved in this scalar product are required to have opposite $SO(5,1)$ chirality, \ie one should be a Weyl spinor and the other an anti-Weyl spinor.

The generators of the superconformal group act on the superspace coordinates $x^{\al\be}$ and $\theta_a^\al$ in the usual way; the resulting transformation laws are (similar transformations but in a different notation, and derived in a different way, also appear in~\cite{Park:1998})
\beqa
\eqnlab{transf_superx}
\de x^{\al \be} & = & a^{\al \be} + \dia{\om}{\ga}{\al} x^{\ga \be} +
\dia{\om}{\ga}{\be} x^{\al \ga} + \la x^{\al \be} +
4 c_{\ga \de} x^{\ga\al} x^{\be\de} - i \Om^{ab}
  \eta^{[\al}_a \theta^{\be]}_b + {} \nn \\
& & {} + c_{\ga \de} \Om^{ac} \Om^{bd} \theta^{[\al}_a
\theta^{\be]}_b \theta^{\ga}_c \theta^{\de}_d - 2i \rho^c_{\ga} \theta^{[\al}_c
    x^{\be]\ga} - \rho^c_{\ga} \Om^{ab} \theta^{[\al}_c
    \theta^{\be]}_a \theta^{\ga}_b \qquad \\
\eqnlab{transf_superth}
\de \theta^{\al}_a & = &(\dia{\om}{\ga}{\al} - 4 c_{\ga \de} x^{\al \de}
- 2i c_{\ga \de} \Om^{cd} \theta_c^{\al} \theta_d^{\de} + 2i
\rho^c_{\ga} \theta^{\al}_c) \theta^{\ga}_a + \frac{1}{2} \la \theta^{\al}_a + {} \nn \\
& & {} + \eta^{\al}_a + 2 \Om_{ca} \rho^c_{\ga} x^{\ga \al} + i
\Om_{ac} \Om^{bd} \rho^c_{\ga} \theta_b^{\ga} \theta^{\al}_d + v_{ac} \Om^{cd}
\theta^{\al}_d,
\eeqa
which, as in the bosonic case, look rather messy and complicated. However, it is
easily shown that the action on the superspace differentials
$e^{\al \be} = d x^{\al \be} + i \Om^{ab} \theta_a^{[\al} d \theta_b^{\be]}$ and $d\theta^\al_a$ can be written as
\beqa
\eqnlab{transf_e}
\de e^{\al \be} & = & \dia{\Om}{\ga}{\al}(x,\theta) e^{\ga \be} +
\dia{\Om}{\ga}{\be}(x,\theta) e^{\al \ga} + \La(x,\theta) e^{\al \be} \\
\eqnlab{transf_dth}
\de (d \theta_a^\al) & = & \dia{\Om}{\ga}{\al}(x,\theta) d \theta_a^\ga +
\frac{1}{2} \La(x,\theta) d\theta_a^\al + V^c_{\ph{d}a}(\theta)
d\theta_c^\al + 2 \Xi_{\ga,a}(\theta) e^{\al \ga},
\eeqa
where the superspace-dependent parameter functions
\beqa
\eqnlab{Omega}
\dia{\Om}{\al}{\ga}(x,\theta) & = & \dia{\om}{\al}{\ga} - 4 c_{\al
  \de} x^{\ga \de} + c \cdot x \kde{\al}{\ga} - 2i c_{\al \de}
\Om^{cd} \theta_c^\ga \theta_d^\de + 2i \rho_{\al}^c \theta^\ga_c -
\frac{i}{2} \dia{\de}{\al}{\ga} \rho \cdot \theta \qquad \\
\eqnlab{Lambda}
\La(x,\theta) & = & \la - 2c \cdot x + i \rho \cdot \theta
\eeqa
are extensions of \Eqsref{Om_bos_al} and \eqnref{La_bos_al} to the
superconformal case, while the functions
\beqa
\eqnlab{V}
V^a_{\ph{a}d}(\theta) & = & - \Om^{ac} v_{cd} + 4i \Om^{ac} c_{\ga \de}
\theta^\ga_c \theta^\de_d -2i \rho^a_\ga \theta^\ga_d + 2i \Om^{ae}
\rho^f_\ga \theta^\ga_e \Om_{fd} \\
\eqnlab{Xi}
\Xi_{\al,a}(\theta) & = & 2 c_{\al \be} \theta^\be_a + \Om_{ab} \rho^b_\al
\eeqa
are new. Naturally, $\dia{\Om}{\al}{\be}$ and $V^a_{\ph{a}b}$ are traceless, \ie they obey $\dia{\Om}{\al}{\al}=V^a_{\ph{a}a}=0$. To the best of our knowledge, this way of presenting the superconformal transformations has not appeared previously in the literature. The advantages of using superspace-dependent parameter functions will be made clearer when we consider transformations of \emph{superfields} in \Secref{superfields}.

The transformations \eqnref{transf_e} and \eqnref{transf_dth} contain
the expected Lorentz, dilatation and $R$-symmetry parts (with
generalized superspace-dependent parameters), but also a term
connecting the variation of $d\theta_a^\al$ to $e^{\al \be}$ with the parameter function $\Xi_{\al,a}(\theta)$, containing the (constant) parameters $c_{\al\be}$ and $\rho^a_\al$. This
separates special conformal and special supersymmetry transformations
from the other transformations, which yield no such possibilities. This
issue will be further discussed later in this paper and is a superspace analogue to the non-trivial piece $k_{\mu}$ in \Eqnref{general_transf}.

It is also interesting to note that the infinitesimal supersymmetric interval length is preserved up to a superspace-dependent scale factor under superconformal transformations, \ie
\beq
\de \left( \frac{1}{2}\eps_{\al\be\ga\de} e^{\al\be} e^{\ga\de} \right) = \La(x,\theta) \eps_{\al\be\ga\de} e^{\al\be} e^{\ga\de}.
\eeq
This relation may in fact be seen as a definition of the superconformal transformations, in analogy with \Eqnref{proptimetransf} in the bosonic case.

Continuing along the path taken in \Secref{bosonic}, we calculate the commutation relations for the differential generators in \Eqsref{super_P}--\eqnref{super_U}. The result is
(similar relations appear \eg in~\cite{Claus:1998,Park:1998})
 \beq
  \eqnlab{super_comm}
  \begin{array}{rclcrcl}
   \com{P_{\al\be}}{M_{\ga}^{\ph{\ga}\de}} & = & -2 \kde{[\al}{\de}
   P_{\be]\ga} - \frac{1}{2} \kde{\ga}{\de} P_{\al\be} & &
   \com{P_{\al \be}}{U^{ab}} & = & 0 \\
   \com{K^{\al\be}}{M_{\ga}^{\ph{\ga}\de}} & = & 2 \kde{\ga}{[\al}
   K^{\be]\de} + \frac{1}{2} \kde{\ga}{\de} K^{\al\be} & &
   \com{K^{\al \be}}{U^{ab}} & = & 0 \\
   \com{U^{ab}}{D} & = & 0 & &
   \com{M_{\al}^{\ph{\al}\be}}{U^{ab}} & = & 0 \\
   \com{U^{ab}}{U^{cd}} & = & - \Om^{a(c} U^{d)b} - \Om^{b(c}
   U^{d)a} & &
   \com{P_{\al \be}}{D} & = & P_{\al \be} \\
   \com{M_{\al}^{\ph{\al}\be}}{D} & = & 0 & &
   \com{K^{\al \be}}{D} & = & - K^{\al \be} \\
   \com{M_{\al}^{\ph{\al}\be}}{Q^a_{\ga}} & = & - \kde{\ga}{\be}
   Q^a_{\al} + \frac{1}{4} \kde{\al}{\be} Q^a_{\ga} & &
   \com{Q_{\al}^a}{D} & = & \frac{1}{2} Q_{\al}^a \\
   \com{\dia{M}{\al}{\be}}{S_a^{\ga}} & = & \kde{\al}{\ga} S_a^{\be} -
   \frac{1}{4} \kde{\al}{\be} S_a^{\ga} & &
   \com{S^{\al}_a}{D} & = & -\frac{1}{2} S^{\al}_a \\
   \com{K^{\al \be}}{Q_{\ga}^a} & = & - 2 \Om^{ac}
   \de_{\ga}^{\ph{\ga}[\al} S_c^{\be]} & &
   \com{P_{\al \be}}{Q_{\ga}^c} & = & 0 \\
   \com{P_{\al \be}}{S^{\ga}_a} & = & 2 \Om_{ac}
   \de_{[\al}^{\ph{[\al}\ga} Q^c_{\be]} & &
   \com{K^{\al \be}}{S^{\ga}_a} & = & 0 \\
   \com{M_{\al}^{\ph{\al}\be}}{M_{\ga}^{\ph{\ga}\de}} & = &
   \kde{\al}{\de} \dia{M}{\ga}{\be} - \kde{\ga}{\be} \dia{M}{\al}{\de}
   & &
   \com{U^{ab}}{Q_{\ga}^c} & = & \Om^{c(a} Q^{b)}_{\ga} \\
   \acom{Q_{\al}^a}{Q_{\be}^b} & = & -2i \Om^{ab} P_{\al \be} & &
   \com{U^{ab}}{S^{\ga}_c} & = & \kde{c}{(a}\Om^{b)d} S_d^{\ga} \\
   \acom{S^{\al}_a}{S^{\be}_b} & = & -2i \Om_{ab} K^{\al \be} \\
   \acom{Q_{\al}^a}{S^{\be}_b} & = & \multicolumn{5}{l}{i
   \kde{\al}{\be} \Big( \de^a_{\ph{a}b} D - 4 \Om_{bc} U^{ac}
   \Big) + 2i \de^a_{\ph{a}b} \dia{M}{\al}{\be}} \\
   \com{P_{\al \be}}{K^{\ga\de}} & = & \multicolumn{5}{l}{ - 4
   \kde{[\al}{[\ga} \dia{M}{\be]}{\de]} - 2 \kde{[\al}{\ga}
   \kde{\be]}{\de} D,}
  \end{array}
 \eeq
which together define the superconformal group $OSp(8^*|4)$.
It should be noted that these are the relations that apply when the differential
operators act on each other, \emph{not} when the generators act on
fields. In the latter case, the sign on the right hand side of every relation should be changed. It is straightforward to verify that the super-Jacobi identities
are satisfied.

In the same way as we compactified the notation in the bosonic case
in~\Eqnref{J_munu} by extending space-time with one extra space-like and one extra time-like dimension, we may define new generators $J_{\alh \beh}$, where
$\alh,\beh = 1,\ldots,8$ are chiral spinor indices in \emph{eight}
dimensions. The hatted indices decompose into subscript and
superscript indices in six dimensions, corresponding to Weyl ($\alh =
1,\ldots,4$) and anti-Weyl ($\alh = 5,\ldots,8$) spinors,
respectively. We let
\beq
  \begin{array}{rcl}
  J_{\al\be} & = & \frac{1}{2} P_{\al\be} \\
  \dia{J}{\al}{\be} & = & \frac{1}{2} \dia{M}{\al}{\be} + \frac{1}{4}
  \kde{\al}{\be} D = - J^\be_{\ph{\be}\al} \\
  J^{\al \be} & = & - \frac{1}{2} K^{\al\be},
  \end{array}
\eeq
and also define new supercharges $\hat{Q}^a_{\alh}$ according to
\beq
  \begin{array}{rcl}
  \hat{Q}^a_\al & = & Q^a_\al \\
  \hat{Q}^{a,\al} & = & \Om^{ab} S^\al_b,
  \end{array}
\eeq
together forming a chiral spinor in eight dimensions.

These generators, together with the unaltered $R$-symmetry generator $U^{ab}$, obey
the commutation relations
 \beq
  \eqnlab{compact_super_comm}
  \begin{array}{rclcrcl}
   \acom{\hat{Q}^a_{\alh}}{\hat{Q}^b_{\beh}} & = & -4i \left( \Om^{ab}
   J_{\alh\beh} + I_{\alh\beh} U^{ab} \right) & \quad & \com{J_{\alh
   \beh}}{U^{ab}} & = & 0 \\[3mm]
   \com{J_{\alh\beh}}{J_{\gah\deh}} & = & - I_{\alh[\gah}
   J_{\deh]\beh} + I_{\beh[\gah} J_{\deh]\alh} & & \com{J_{\alh
   \beh}}{\hat{Q}^a_{\gah}} & = & I_{\gah[\alh} \hat{Q}^a_{\beh]} \\[3mm]
   \com{U^{ab}}{U^{cd}} & = & - \Om^{a(c} U^{d)b} - \Om^{b(c} U^{d)a}
   & & \com{U^{ab}}{\hat{Q}^c_{\alh}} & = & \Om^{c(a}
   \hat{Q}^{b)}_{\alh}, \\
  \end{array}
 \eeq
where the symmetric matrix $I_{\alh \beh}$ has components
 \beq
  \begin{array}{rcl}
  I_{\al\be} & = & 0 \\
  \dia{I}{\al}{\be} & = & \kde{\al}{\be} = I^\be_{\ph{\be}\al} \\
  I^{\al \be} & = & 0;
  \end{array}
 \eeq
it transforms in the singlet representation of $SO(6,2)$. The
commutation relations \eqnref{compact_super_comm} contain all the
information of \Eqnref{super_comm}, but in a much more compact
notation.

It is possible to compactify the notation further by considering the
generators in a superspace with eight bosonic and four fermionic
dimensions. This yields the $OSp(8^*|4)$ structure in a manifest way;
this notation appears in \cite{Kac:1977,Claus:1998}. Introduce a
matrix $J_{\sss AB}$, where $A=(\alh,a)$ and $B=(\beh,b)$ are
$OSp(8^*|4)$ indices. Note that $A$ and $B$ are superindices,
meaning that $J_{\sss AB}$ is symmetric if both indices are fermionic,
otherwise it is antisymmetric. We denote this in the standard way by
\beq
J_{\sss AB} = - (-1)^{\sss AB} J_{\sss BA}.
\eeq
$J_{\sss AB}$ contains the different generators of the superconformal
group, explicitly we take
\beq
  \begin{array}{rcl}
  J_{\alh \beh} & = & J_{\alh \beh} \\
  \dia{J}{\alh}{b} & = & \frac{i}{2\sqrt{2}} \hat{Q}^b_{\alh} \\
  J^a_{\ph{a}\beh} & = & - \frac{i}{2\sqrt{2}} \hat{Q}^a_{\beh} \\
  J^{ab} & = & i U^{ab}.
  \end{array}
\eeq
These generators obey the (anti)commutation relations
\beq
\eqnlab{superalgebra}
\Big[J_{\sss AB}, J_{\sss CD} \Big\} = -\frac{1}{2} \Big( I_{\sss BC}
J_{\sss AD} - (-1)^{\sss AB} I_{\sss AC} J_{\sss BD} -
(-1)^{\sss CD} I_{\sss BD} J_{\sss AC} +
(-1)^{\sss AB+CD} I_{\sss AD} J_{\sss BC} \Big)
\eeq
where the bracket in the left hand side is an anticommutator if both
entries in it are fermionic, otherwise it is a commutator.
The superspace metric $I_{\sss AB}=(-1)^{\sss AB} I_{\sss BA}$ is defined by
\beq
  I_{\sss AB} =
  \left( \begin{array}{ccc}
  0 &  \kde{\al}{\be} & 0 \\
  \de^\al_{\ph{\al}\be} & 0 & 0 \\
  0 & 0 & i \Om^{ab}
  \end{array} \right).
  \eqnlab{superspacemetric}
\eeq
In order to make the relation
\beq
I_{\sss AB} I^{\sss BC} = \de_{\sss A}^{\sss C}
\eeq
valid (which is essential if want to raise and lower indices), we also need to define the inverse superspace metric as
\beq
  I^{\sss AB} =
  \left( \begin{array}{ccc}
  0 &  \de^\al_{\ph{\al}\be} & 0 \\
  \kde{\al}{\be} & 0 & 0 \\
  0 & 0 & -i \Om_{ab}
  \end{array} \right).
  \eqnlab{superspacemetric_upper}
\eeq
Note the resemblance between the (anti)commutation relations in
\Eqnref{superalgebra} and the well-known Lorentz group commutation
relations. This suggests that the superconformal transformations act linearly in a superspace with eight bosonic and four fermionic dimensions.

In \Secref{conf_bos}, we found the conformal transformations of the bosonic coordinates $x^\mu$ in an indirect way, by looking at a projective hypercone embedded in an eight-dimensional space with two time-like directions. In this
higher-dimensional space, the conformal group acts linearly. Guided by
the $OSp(8^*|4)$ covariant notation introduced above, we would like to
perform a similar analysis in the superconformal case.

Let the coordinates in superspace be $y_{\sss A}$ and introduce a projective
supercone by the equation
\beq
\eqnlab{supercone}
I^{\sss AB} y_{\sss A} y_{\sss B} = 0.
\eeq
We will parametrize the supercone in a more implicit manner than we did when considering the hypercone in
\Secref{conf_bos}. Consider a point on the supercone, with coordinates
$y_{\sss A}=(y_\al,y^\al,y^a)$. It is always possible to introduce a
fermionic field $\theta_a^\al(y)$ such that for any point on the supercone,
\beq
\eqnlab{y^a}
y^a = \sqrt{2} \Om^{ab} \theta_b^\be y_\be.
\eeq
By requiring $\theta_a^\al$ to transform as an anti-Weyl spinor under SO(5,1),
this field is well-defined in all points on the supercone. In the same manner,
we introduce the bosonic field $x^{\al\be}(y)=-x^{\be\al}(y)$ such that
\beq
\eqnlab{y^al}
y^\al = \left( 2 x^{\al\be} - i \Om^{ab} \theta^\al_a \theta^\be_b \right) y_\be
\eeq
for any point on the supercone.

It is easily verified that all points $y_{\sss A}$ of this form lie on the
supercone defined by \Eqnref{supercone}. Obviously, we may always multiply
$x^{\al\be}$ or $\theta_a^\al$ by a constant and still remain on the supercone.
This explains the notion \emph{projective} supercone.

The next step is to vary these coordinates. The transformations are generated by
\beq
J_{\sss AB} = - y_{\sss [A} \pa_{\sss B]} = (-1)^{\sss AB} y_{\sss [B} \pa_{\sss A]},
\eeq
which satisfies the commutation relations~\eqnref{superalgebra}, given that
\beq
\pa_{\sss A} y_{\sss B} = I_{\sss AB}.
\eeq
The coordinates $y_{\sss A}$ transform according to
\beq
\de y_{\sss A} = \pi^{\sss CD} J_{\sss CD} y_{\sss A} = (-1)^{\sss C(A+D)} I_{\sss AC} \pi^{\sss CD} y_{\sss D},
\eeq
where the parameter matrix is
given by
\beq
  \pi^{\sss AB} =
  \left( \begin{array}{ccc}
  2 a^{\al\be} & \dia{\om}{\be}{\al} + \frac{1}{2} \la \kde{\be}{\al} &
    -i \sqrt{2} \eta^\al_b \\
  - \dia{\om}{\al}{\be} - \frac{1}{2} \la \kde{\al}{\be} &
    -2 c_{\al\be} & -i \sqrt{2} \rho^c_\al \Om_{cb} \\
  i\sqrt{2} \eta^\be_a & i\sqrt{2} \rho^c_\be \Om_{ca} & -i v_{ab}
  \end{array} \right),
\eeq
chosen such that the relation
\beq
\pi^{\sss AB} J_{\sss AB} = \dia{\om}{\al}{\be} \dia{M}{\be}{\al} + a^{\al\be} P_{\al\be}
+ c_{\al\be} K^{\al\be} + \la D + \eta^\al_a Q^a_\al + \rho^a_\al S_a^\al + v_{ab} U^{ab}
\eeq
is valid.

Since \Eqnref{supercone} is invariant under a transformation of this type, we may require the left-hand and the right-hand sides of \Eqsref{y^a}--\eqnref{y^al} to transform equally.
The implicated transformation properties of the fields $x^{\al\be}(y)$ and $\theta_a^\al(y)$ when the $y$-coordinates are transformed in this way are found to agree exactly with the superconformal transformations of the coordinates $x^{\al\be}$ and $\theta_a^\al$ in our original superspace (with six bosonic and sixteen fermionic dimensions) in
\Eqsref{transf_superx} and \eqnref{transf_superth} above! This explains the
choice of notation and implies that the rather complicated transformation laws for
$x^{\al\be}$ and $\theta_a^\al$ are a mere consequence of a simple rotation
in a superspace with eight bosonic and four fermionic dimensions!

This way of introducing the superspace coordinates and deriving their transformation properties has, as far as we know, not appeared previously in the literature. Presumably, the fact that this works points to some underlying structure of the $(2,0)$ superspace in six dimensions, the nature of which is not clear at the moment.

\subsection{Superfields}
\seclab{superfields}

Having introduced a superspace (in the remainder of this paper, we will work in the usual $(2,0)$ superspace with six bosonic and sixteen fermionic dimensions), the next step is to populate it with
superfields. The superfield formulation for the $(2,0)$ tensor
multiplet first appeared in~\cite{Howe:1983}; a thorough description
of its use in our model can be found in~\cite{AFH:2003coupled}. In
this paper, we will content ourselves with a short description of the
key aspects.

Define a superfield $\Phi^{ab}=\Phi^{ab}(x,\theta)$, obeying the
algebraic constraint
\beq
\Om_{ab} \Phi^{ab} = 0
\eeq
and the differential constraint
\beq
\eqnlab{diff_con}
D^a_{\al} \Phi^{bc} + \frac{2}{5} \Om_{de} D^d_\al \left( \Om^{ab}
\Phi^{ec} + \Om^{ca} \Phi^{eb} + \frac{1}{2} \Om^{bc} \Phi^{ea}
\right) = 0,
\eeq
where $D^a_{\al}$ is the covariant superspace derivative, defined
according to
\beq
D^a_{\al}  = \pa^a_{\al} + i \Om^{ac} \theta_c^{\ga} \pa_{\al\ga}.
\eeq
It is important to note~\cite{AFH:2003coupled} that the differential
constraint \eqnref{diff_con} implies that the lowest component of the superfield must obey the free equations of motion for a massless scalar field, \ie the Klein-Gordon equation. This reflects the fact
that we are dealing with an \emph{on-shell} superfield formulation.

It is convenient to define supplementary superfields according to
\beqa
\Psi^a_{\al} & = & - \frac{2i}{5} \Om_{bc} D^b_\al \Phi^{ca} \\
H_{\al \be} & = & \frac{1}{4} \Om_{ab} D^a_{\al} \Psi^b_{\be},
\eeqa
but it should be noted that these contain no new degrees of freedom compared to
$\Phi^{ab}$.

By definition, a superfield transforms according to
\beq
\de_{\sss Q} \Phi^{ab} = \com{\eta \cdot Q}{\Phi^{ab}}
\eeq
under a supersymmetry transformation. Working out this commutator,
with $Q$ as in \Eqnref{Q_super}, it can be shown (by comparing with
the explicit transformations in~\cite{AFH:2003free}) that the lowest
components of the superfields $\Phi^{ab}$, $\Psi^a_{\al}$ and
$H_{\al\be}$ are the tensor multiplet fields $\phi^{ab}$, $\psi^a_\al$
and $h_{\al\be}$, hence the choice of notation. The differential
constraint~\eqnref{diff_con} yields the usual free equations of motion for
these component fields.

The purpose of this section is to find how the rest of the superconformal transformations
act on the superfields. The transformations will, as in the bosonic
case, contain one piece including the differential expressions in
\Eqsref{super_P}--\eqnref{super_U} and some non-differential pieces. The
latter may be derived by requiring that the transformation of $\Phi^{ab}$ must
satisfy the differential constraint~\eqnref{diff_con} when $\Phi^{ab}$
itself does. We also require the abstract transformation operator to
commute with the covariant derivative in superspace (in the same way
as transformations commute with derivatives in the bosonic case, see
\Eqnref{bos_der_comm} and the discussion thereafter). Note that this approach
of course requires the superfields to be on shell. The transformation
of $(2,0)$ superfields was also discussed in~\cite{Grojean:1998} using a
geometric approach, realizing the transformations as derivations in superspace. We will be more explicit and algebraic in our treatment of the problem, trying to take advantage of the superfield
formulation. It is our goal to write the transformations of the
superfields in a form similar to the one used in the bosonic case in
\eg \Eqnref{transf_phi_al}, inspired by the transformation properties of the superspace differentials in \Eqsref{transf_e}--\eqnref{transf_dth}.

After some quite involved computations, it is found that the
superfields transform according to (we are still in the active
picture, where we transform the fields rather than the coordinates)
\beqa
\eqnlab{dephi}
\de_{\sss C} \Phi^{ab} & = & \xi^{\ga \de}(x,\theta) \pa_{\ga \de} \Phi^{ab} +
\xi_c^{\ga}(x,\theta) \pa^c_\ga \Phi^{ab} + 2 \La(x,\theta) \Phi^{ab} + {} \nn \\ & & {} + V^a_{\ph{a}d}(\theta) \Phi^{db} + V^b_{\ph{b}d}(\theta) \Phi^{ad} \\
\eqnlab{depsi}
\de_{\sss C} \Psi^a_{\al} & = & \xi^{\ga \de}(x,\theta) \pa_{\ga \de} \Psi^a_{\al} +
\xi_c^{\ga}(x,\theta) \pa^c_\ga \Psi^a_{\al} + \dia{\Om}{\al}{\ga}(x,\theta) \Psi^a_\ga + {} \nn \\
& & {} + \frac{5}{2} \La(x,\theta) \Psi^a_{\al} + V^a_{\ph{a}c}(\theta) \Psi^c_{\al} - 4 \Xi_{\al,b}(\theta) \Phi^{ab} \\
\eqnlab{deh}
\de_{\sss C} H_{\al \be} & = & \xi^{\ga \de}(x,\theta) \pa_{\ga \de} H_{\al\be} +
\xi_c^{\ga}(x,\theta) \pa^c_\ga H_{\al\be} + \dia{\Om}{\al}{\ga}(x,\theta) H_{\ga\be} +
\dia{\Om}{\be}{\ga}(x,\theta) H_{\al\ga} + {} \nn \\
& & {} + 3 \La(x,\theta) H_{\al \be} + 3i \Xi_{\al,c}(\theta) \Psi^c_\be + 3i \Xi_{\be,c}(\theta) \Psi^c_\al,
\eeqa
where the parameter functions $\dia{\Om}{\al}{\be}(x,\theta)$, $\La(x,\theta)$, $V^a_{\ph{a}b}(\theta)$ and $\Xi_{\al,a}(\theta)$
are those defined in \Eqsref{Omega}--\eqnref{Xi}, respectively, while $\xi^{\al
  \be}(x,\theta) = \de x^{\al \be}$ and $\xi_a^{\al}(x,\theta)= \de \theta_a^{\al}$, see
\Eqsref{transf_superx} and \eqnref{transf_superth}. In analogy with the notion of primary fields in the
purely bosonic case, we see that the superfield $\Phi^{ab}$ is \emph{superprimary}
(its transformation does not contain any $\Xi$-part) while the others
are not. Note that the transformations (in this notation) are what one would expect by looking at the indices and mass dimensions of the fields, apart from
the parts containing $\Xi_{\al,a}(\theta)$, where the numerical
coefficients are hard to guess a priori.

From these transformations, the transformation laws for the component
fields may be read off. The $SO(6,2)$ transformations agree, as expected, with
\Eqsref{transf_psi}, \eqnref{transf_phi_al} and \eqnref{transf_h_al},
while supersymmetry acts according to
 \beq
  \begin{array}{rcl}
\de_{\sss Q} \phi^{ab} & = & i \eta^{\al}_c \left( \Om^{ac}
\psi^b_{\al} - \Om^{bc} \psi^a_{\al} - \frac{1}{2}
\Om^{ab} \psi^c_{\al} \right) \\[1mm]
\de_{\sss Q} \psi^a_{\al} & = & \Om^{ab} \eta_b^{\be} h_{\al \be} + 2 \pa_{\al
  \be} \phi^{ab} \eta_b^{\be} \\[1mm]
\de_{\sss Q} h_{\al \be} & = & i \eta_a^{\ga} \left( \pa_{\ga \al}
\psi^a_{\be} + \pa_{\ga \be} \psi^a_{\al} \right),
  \end{array}
 \eeq
in agreement with~\cite{AFH:2003free}.
The special supersymmetry transformations of the component fields are
 \beq
  \begin{array}{rcl}
\de_{\sss S} \phi^{ab} & = & -4i x^{\ga \de} \rho_\ga^{[a} \psi^{b]}_\de - i
\Om^{ab} \Om_{cd} x^{\ga \de} \rho_\ga^c \psi^d_\de \\[1mm]
\de_{\sss S} \psi^a_{\al} & = & -2 \rho^a_\ga x^{\ga\de} h_{\de \al} - 4
\rho^b_\al \Om_{bc} \phi^{ca} + 4 \rho^b_\ga \Om_{bc} x^{\ga \de}
\pa_{\al \de} \phi^{ac} \\[1mm]
\de_{\sss S} h_{\al \be} & = & 2i \Om_{ab} \rho^a_\ga x^{\ga \de} \Big(
\pa_{\de \al} \psi_\be^b + \pa_{\de \be} \psi_\al^b \Big) -3i \Om_{ab}
\Big( \rho^a_\al \psi^b_\be + \rho^a_\be \psi^b_\al \Big),
  \end{array}
 \eeq
while $R$-symmetry acts according to
 \beq
  \begin{array}{rcl}
\de_{\sss U} \phi^{ab} & = & - \Om^{ac} v_{cd} \phi^{db} - \Om^{bc}
v_{cd} \phi^{ad} \\[1mm]
\de_{\sss U} \psi^{a}_\al & = & - \Om^{ac} v_{cd} \psi^{d}_\al \\[1mm]
\de_{\sss U} h_{\al \be} & = & 0,
  \end{array}
 \eeq
as suggested by the index structure of the fields.

This completes the analysis of the superconformal transformations of the free tensor multiplet in
superspace, but we should also consider the self-dual string. It is,
as before, described by a bosonic embedding field $X^{\al \be}(\si)$, but
we have to supplement it by a second (fermionic) field $\Theta^\al_a(\si)$
describing the embedding in the fermionic coordinates. The
superconformal transformations of these fields are
\beqa
\eqnlab{transf_superX}
\de_{\sss C} X^{\al \be} & = & - a^{\al \be} - \dia{\om}{\ga}{\al} X^{\ga \be} -
\dia{\om}{\ga}{\be} X^{\al \ga} - \la X^{\al \be} -
4 c_{\ga \de} X^{\ga\al} X^{\be\de} + i \Om^{ab}
  \eta^{[\al}_a \Theta^{\be]}_b - {} \nn \\
& & {} - c_{\ga \de} \Om^{ac} \Om^{bd} \Theta^{[\al}_a
\Theta^{\be]}_b \Theta^{\ga}_c \Theta^{\de}_d + 2i \rho^c_{\ga} \Theta^{[\al}_c
    X^{\be]\ga} + \rho^c_{\ga} \Om^{ab} \Theta^{[\al}_c
    \Theta^{\be]}_a \Theta^{\ga}_b \qquad \\
\eqnlab{transf_superTh}
\de_{\sss C} \Theta^{\al}_a & = & -(\dia{\om}{\ga}{\al} - 4 c_{\ga \de} X^{\al \de}
- 2i c_{\ga \de} \Om^{cd} \Theta_c^{\al} \Theta_d^{\de} + 2i
\rho^c_{\ga} \Theta^{\al}_c) \Theta^{\ga}_a - \frac{1}{2} \la \Theta^{\al}_a - {} \nn \\
& & {} - \eta^{\al}_a - 2 \Om_{ca} \rho^c_{\ga} X^{\ga \al} - i
\Om_{ac} \Om^{bd} \rho^c_{\ga} \Theta_b^{\ga} \Theta^{\al}_d - v_{ac} \Om^{cd}
\Theta^{\al}_d,
\eeqa
which coincide with
\Eqsref{transf_superx} and \eqnref{transf_superth} for the
corresponding coordinates but with the \emph{opposite sign}, in the
same manner as in the purely bosonic model. These variations do not have the same structure as the variations of the superfields $\Phi^{ab}$, $\Psi^a_\al$ and $H_{\al\be}$, but if we instead consider the appropriate differentials of these fields, we get something more familiar. The differentials in question are related in an obvious way to the superfield differentials $e^{\al\be}$ and $d\theta^\al_a$ and are $E^{\al\be} = \tilde{d} X^{\al\be} + i \Om^{ab} \Th_a^{[\al} \tilde{d} \Th_b^{\be]}$ and $\tilde{d} \Theta^\al_a$, where $\tilde{d}=d\si^i \pa_i$ denotes a differential operator with respect to the world-sheet parameters $\si$. The variations of these quantities are
\beqa
\eqnlab{transf_E}
\de_{\sss C} E^{\al \be} & = & - \dia{\Om}{\ga}{\al}(X,\Theta) E^{\ga \be} -
\dia{\Om}{\ga}{\be}(X,\Theta) E^{\al \ga} - \La(X,\Theta) E^{\al \be} \\
\eqnlab{transf_dTh}
\de_{\sss C} (\tilde{d} \Theta_a^\al) & = & - \dia{\Om}{\ga}{\al}(X,\Theta) \tilde{d} \Theta_a^\ga -
\frac{1}{2} \La(X,\Theta) \tilde{d} \Theta_a^\al - V^d_{\ph{d}a}(\Theta)
\tilde{d}\Theta_d^\al - 2 \Xi_{\ga,a}(\Theta) E^{\al \ga}, \qquad \quad
\eeqa
which are similar to \Eqsref{transf_e} and \eqnref{transf_dth} but again with the opposite sign. We see that $E^{\al\be}$ transforms as a superprimary quantity while $\tilde{d} \Theta^\al_a$ does not.

In retrospect, what we have done in this section is to generalize the
superfield formulation, thereby incorporating the full superconformal
group in the formalism. This is a must in order to calculate
variations of terms involving superfields under such transformations;
using the transformation laws for the component fields would lead us
nowhere (we do not even know the explicit expressions for the superfields in
terms of component fields to all orders). With the superspace
generators, we may in a relatively simple and compact way describe the
transformations.

We should also comment on the use of superspace-dependent parameter functions in the transformation laws. This also facilitates the calculations, since Lorentz and $R$-symmetry covariance is manifest through the use of tensor notation. They must, however, be applied with care since these parameter functions generally do not commute with derivatives. To the best of our knowledge, this approach to the superconformal transformations of $(2,0)$ superfields has not appeared previously in the literature.

It would be interesting to consider the problem of finding a field theory in the superspace with eight bosonic and four fermionic dimensions discussed at the end of the previous subsection. This theory would, presumably, incorporate manifest superconformal symmetry and probably yield some new insights into the properties of the six-dimensional theory.

\subsection{The supersymmetric interaction terms}
\seclab{Wess-Zumino}

In this subsection, we consider the theory describing the supersymmetric coupling of a self-dual string to a $(2,0)$ tensor multiplet background~\cite{AFH:2003coupled}. As usual, background coupling means that the
tensor multiplet fields are taken to be on shell, \ie they obey their free equations of motion.

The interaction is described by the action terms
\beq
\eqnlab{interaction}
S^{\mathrm{int}} = - \int_{\Si} d^2 \si \sqrt{\Phi \cdot \Phi}
\sqrt{-G} + \int_D F,
\eeq
where the first term is a Nambu-Goto term where the string tension has been replaced by an expression in the superfield $\Phi^{ab}$ discussed in the previous subsection. Explicitly, the $SO(5)$ invariant scalar product is defined by
\beq
\eqnlab{so5scalar}
\Phi \cdot \Phi \equiv \frac{1}{4} \Om_{ac} \Om_{bd} \Phi^{ab}
\Phi^{cd},
\eeq
and its appearance can be understood by considering the relation between the moduli parameters and the string tension mentioned in the introduction. Note that the term involves the \emph{pull-back} of the superfield $\Phi^{ab}$ to the string world-sheet $\Si$. In the second factor of the Nambu-Goto term, $G$ denotes the determinant of the induced metric in superspace, which may be written as
\beq
G = \frac{1}{2} \eps^{ik} \eps^{jl} G_{ij} G_{kl}
\eeq
where
\beq
G_{ij} = \frac{1}{2} \eps_{\al\be\ga\de} E^{\al\be}_i E^{\ga\de}_j.
\eeq
Here $E^{\al\be}_i$ is determined by the relation
$E^{\al\be} = d \si^i E_i^{\al\be}$. Naturally, $\Si$ denotes the string world-sheet.

The second term in \Eqnref{interaction} is of Wess-Zumino type and involves the pull-back of a certain super three-form $F$ to the world-volume of a "Dirac membrane" $D$ attached to the string, satisfying $\pa D = \Si$. For a
more elaborate discussion of the Dirac membrane and its properties, we
refer to~\cite{AFH:2003coupled,Deser:1998}. $D$ is described, similarly to the string,
by embedding fields $X^{\al\be}$ and $\Theta^\al_a$, but these fields are functions of three parameters instead of two. As an embedding, they naturally transform in the same way as the string embedding fields, \ie according to
\Eqsref{transf_superX} and \eqnref{transf_superTh}, under a superconformal transformation.

The super three-form $F$ is expressed as
\beqa
F & = & \frac{1}{6} e^{\ga_1 \ga_2} \we e^{\be_1 \be_2} \we e^{\al_1 \al_2} \eps_{\al_2 \be_2 \ga_1 \ga_2} H_{\al_1 \be_1} - \frac{i}{2} e^{\ga_1 \ga_2} \we e^{\be_1 \be_2} \we d \theta^\al_a \eps_{\al \be_1 \be_2 \ga_1} \Psi_{\ga_2}^a + {} \nn \\
\eqnlab{F}
& & {} + \frac{i}{2} e^{\ga_1 \ga_2} \we d \theta^\be_b \we d \theta^\al_a \eps_{\al \be \ga_1 \ga_2} \Phi^{ab}
\eeqa
in terms of the superfields $\Phi^{ab}$, $\Psi^a_\al$ and $H_{\al\be}$. It was used by us in~\cite{AFH:2003coupled} but has also appeared (in a slightly different notation) in~\cite{Grojean:1998,Howe:2000}. The coefficients in the expression for $F$ make it a closed form in superspace, \ie $dF=0$, where $d$ is the superspace exterior derivative. The last statement is only true if the superfields obey the differential constraint \eqnref{diff_con}, but this is always the case on shell.

The relative coefficient in \Eqnref{interaction} is determined by requiring invariance under a local fermionic $\kappa$-symmetry, which also removes half of the degrees of freedom contained in $\Theta^\al_a$, as required by
supersymmetry~\cite{AFH:2003coupled}. The closure of $F$ is essential
for this to work, but it also implies that the choice
of Dirac membrane $D$, given a string world-sheet $\Si=\pa D$,
should have no physical significance.

Next, let us consider a superconformal variation of the interaction \eqnref{interaction}. The Nambu-Goto term is clearly invariant; this is easily seen from the expressions \eqnref{dephi} and \eqnref{transf_E} for the variations of $\Phi^{ab}$ and $E^{\al\be}$, respectively. Since all fields are superprimary, the only thing we really need to check is that the conformal weights match appropriately.

The variation of the Wess-Zumino term is a bit
more involved, mainly because of the terms involving
$\Xi_{\al,a}(\theta)$ in the transformation laws for
$\tilde{d} \Theta^{\al}_a$, $\Psi^a_\al$ and $H_{\al\be}$. It turns out that
the pull-back of the super three-form $F$ to the Dirac membrane world-volume $D$ is superconformally
invariant if and only if the coefficients are chosen as in \Eqnref{F},
\ie with the same choice of coefficients that makes it a
closed three-form in superspace! This means that the interaction in
\Eqnref{interaction} indeed is superconformally invariant, but \emph{only} with
the specific three-form in \Eqnref{F}! It should also be stressed that, out of all the transformations in the superconformal group,
it is only the special conformal and the special supersymmetry
transformations that put any restrictions on the coefficients. With
another choice, the theory would still be \eg supersymmetric. The
model is therefore an example of a theory where special conformal
invariance does not follow immediately from dilatational,
translational and Lorentz invariance.

Summing up, we have found that the requirement of $\kappa$-symmetry
and that of superconformal invariance impose equivalent restrictions on
the coefficients of the super three-form $F$, but in different
ways. Note, however, that $\kappa$-symmetry also determines the
relative coefficient between the Nambu-Goto term and the Wess-Zumino
term. The superconformal symmetry of the theory has no influence on
that. The remarkable agreement between superconformal invariance and
$\kappa$-symmetry indicates the uniqueness of the theory: the model is
tightly constrained by its symmetries. This is arguably the most important result of this paper.

\section{The Poincar\'e dual to the string world-sheet}
\seclab{poincare}

In the previous section, we worked with an on-shell tensor multiplet. This implied that the super three-form $F$ in \Eqnref{F} was closed and lead us to the conclusion that the choice of Dirac membrane, given a specific string world-sheet $\Si$, should have no physical significance.

In this section, we try to relax the requirement that $F$ should be closed, as a step in the process of finding a theory where the tensor multiplet fields need not be on shell. We will start in the bosonic case, where the motivation for these ideas is found, and then turn to the superconformal model and try to generalize the concepts introduced in the bosonic theory.

\subsection{The bosonic model}

A free two-form gauge field $b$ with a three-form field strength $h$ obeys the Bianchi identity
\beq
dh=0
\eeq
and the equations of motion
\beq
d*h=0,
\eeq
where $*$ denotes the Hodge duality operator. For a self-dual three-form $h$, these two equations coincide.

If we want to couple this field electromagnetically to a self-dual string (with both magnetic and electric charge), this is done by changing the right-hand sides of the equations above. Explicitly, we get
\beq
  \eqnlab{dh=desi}
  \begin{array}{rcl}
    dh & = & \desi \\
    d*h & = & \desi,
  \end{array}
\eeq
where $\desi$ denotes a four-form called the Poincar\'e dual to the string world-sheet $\Si$. It is defined by the relation
\beq
\int_\Si b \equiv \int_M \desi \we b,
\eeq
where the left integral is evaluated over the string world-sheet, while the right is over the entire six-dimensional Minkowski space. We also note that the self-dual piece of $h$, expressed as $h_+=\frac{1}{2}(h+*h)$, obeys an inhomogeneous equation while the anti self-dual piece $h_-=\frac{1}{2}(h-*h)$ is free and therefore decoupled from the theory, as required. It should be emphasized that \Eqnref{dh=desi} does not \emph{imply} self-duality, but it is \emph{consistent} with self-duality.

Explicitly, the Poincar\'e dual is expressed in terms of the string embedding fields $X^{\mu}$ and the space-time coordinates $x^{\mu}$ according to
\beq
\eqnlab{desi}
\desi = \frac{1}{4!} dx^\mu \we dx^\nu \we dx^\rho \we dx^\si \int_\Si \tilde{d}X^\tau  \we \tilde{d}X^\veps \eps_{\mu\nu\rho\si\tau\veps} \de^{(6)}(x-X),
\eeq
where $\tilde{d} \equiv d\si^i \pa_i$ again is the differential with respect to the world-sheet variables $\si^i$, $i=1,2$, and $\de^{(6)}(x-X)$ is a Dirac delta function in six dimensions. This is somewhat analogous to the coupling of a dyonic relativistic particle to a Maxwell field in four dimensions, in that sense $\desi$ is a generalization of the current four-vector~\cite{Deser:1998}. In our model, we couple to a string and $\desi$ is the dual of a current two-form.

In order for this four-form to be consistent with \Eqnref{dh=desi}, it must be closed. Applying the exterior derivative to $\desi$, we get that
\beqa
d \desi & = & \frac{1}{4!} dx^\mu \we dx^\nu \we dx^\rho \we dx^\si \we dx^{\ka} \int_\Si \tilde{d}X^\tau  \we \tilde{d}X^\veps \eps_{\mu\nu\rho\si\tau\veps}  \pa_{\ka} \de^{(6)}(x-X) = \nn \\
& = & - \frac{2}{5!} dx^\mu \we dx^\nu \we dx^\rho \we dx^\si \we dx^\ka \int_\Si \tilde{d} X^{\tau} \we \tilde{d} X^\veps \eps_{\ka\mu\nu\rho\si\tau} \pa_{\veps} \de^{(6)}(x-X) = \nn \\
\eqnlab{ddesi=0}
& = & - \frac{2}{5!} dx^\mu \we dx^\nu \we dx^\rho \we dx^\si \we dx^\ka \int_\Si \tilde{d} \left( \tilde{d}X^{\tau} \eps_{\ka\mu\nu\rho\si\tau} \de^{(6)}(x-X) \right) = 0, \quad \quad
\eeqa
where we used the fact that an antisymmetrization over seven vector indices in six dimensions always is zero. The last equality follows since the integrand is a total derivative and the string world-sheet $\Si$ has no boundary.

Next, we are interested in how these relations and quantities behave under (bosonic) conformal transformations. Consider the variation of the three-form $h$, which according to \Eqnref{transf_h} equals
\beq
\de_{\sss C} h = \frac{1}{3!} dx^{\mu} \we dx^{\nu} \we dx^{\rho} \Big( \xi^{\si}(x) \pa_{\si} h_{\mu\nu\rho}(x) - 3 \dia{\Om}{\mu}{\si}(x) h_{\si \nu \rho}(x) + 3 \La(x) h_{\mu\nu\rho}(x) \Big).
\eeq
However, if we compare this with \Eqnref{transf_dx}, we see that
\beq
\eqnlab{act_pas_bos}
\de_{\sss C} h = \de_{x} h,
\eeq
where $\de_{x}$ denotes a \emph{passive} conformal transformation, \ie one that acts on the space-time coordinates rather than on the fields. This means that, when dealing with $h$ as a differential form, passive and active transformations yield the same result! We will denote such forms as \emph{primary} differential forms. The statement remains true also when exterior derivatives are applied, \ie $\de_{\sss C} (dh) = \de_{x} (dh)$, but also for the Hodge dual of $h$, which means that $\de_{\sss C} (*h) = \de_{x} (*h)$.

Considering \Eqnref{dh=desi}, we note that the left-hand sides transform as primary four-forms, therefore the right-hand sides should transform in the same way. Since $\desi$ is an expression in terms of the embedding field $X^{\mu}$, we may calculate its active transformation explicitly. Before doing this, we want to rewrite the transformation of $\tilde{d}X^\mu$ in \Eqnref{transf_dX} so that all parameter functions are expressed in terms of the space-time coordinates $x^\mu$, rather than in terms of the embedding field $X^{\mu}$. We find that
\beqa
\de_{\sss C} \Big( \tilde{d}X^\mu \Big) & = & - \Om^{\mu}_{\ph{\mu}\nu}(x) \tilde{d}X^\nu - \La(x) \tilde{d}X^\mu - 2 c_\nu \tilde{d} \Big( (x-X)^\mu (x-X)^\nu \Big) + {} \nn \\
& & {} + c^\mu \tilde{d} \Big( (x-X) \cdot (x-X) \Big),
\eeqa
while the passive variation of this quantity obviously vanishes. We will also need the relation
\beq
\de_{\sss C} \Big( d x^\mu \Big) = 0 = \de_x dx^{\mu} - \Om^{\mu}_{\ph{\mu}\nu}(x) dx^\nu - \La(x) dx^\mu.
\eeq
Finally, considering the Dirac delta function we find that
\beq
\de_{\sss C} \left( \de^{(6)}(x-X) \right) = - \de_{\sss C} X^{\mu} \pa_{\mu} \de^{(6)}(x-X) = \de_x \left( \de^{(6)}(x-X) \right) + 6 \La(x) \de^{(6)}(x-X),
\eeq
where we have used the relation
\beq
\eqnlab{delta_x}
x^\mu \pa_{\nu} \de^{(6)}(x) = - \kde{\nu}{\mu} \de^{(6)}(x).
\eeq
Putting all this together, again using the properties of the Dirac delta function, we arrive at the conclusion
\beq
\eqnlab{de=de}
\de_{\sss C} (\desi) = \de_x (\desi),
\eeq
meaning that our expression \eqnref{desi} for $\desi$ transforms as a primary four-form. This shows that \Eqnref{dh=desi} is a well-defined equation with respect to conformal symmetry.

\subsection{The superconformal model}

Having investigated the bosonic case, we try to generalize these concepts to superspace. We note that the on-shell super three-form $F$ in~\Eqnref{F} is primary, \ie
\beq
\de_{\sss C} F = \de_{x,\theta} F,
\eeq
where $\de_{x,\theta}$ obviously denotes a passive conformal transformation in superspace. The validity of this equation is most easily seen from the fact that $F$ yields a superconformally invariant Wess-Zumino term, see \Secref{Wess-Zumino}.

Let us now try to take $F$ off shell, \ie relax the requirement $dF=0$. The equation corresponding to \Eqnref{dh=desi} is
\beq
\eqnlab{calF}
d \calF = \De_{\sss \Si},
\eeq
where $d$ is the exterior derivative in superspace and $\calF$ is a super three-form, not necessarily equal to $F$ defined in \Eqnref{F}.
The super four-form $\De_{\sss \Si}$ appearing on the right-hand side is supposed to be a generalization of the Poincar\'e dual $\desi$. To find such a quantity is non-trivial, since there is no proper analogue of the Poincar\'e dual in superspace. The best we can hope for is to find a super four-form that reduces to $\desi$ if all fermionic degrees of freedom are removed, and that transforms in a nice way under superconformal transformations.

Guided by our previous experience, we try to formulate this four-form in terms of superfields. The fundamental superfields involving the embedding fields $X^{\al\be}$ and $\Theta^\al_a$ are
\beqa
s^{\al\be} & \equiv & x^{\al\be} - X^{\al\be} - i \Om^{ab} \theta^{[\al}_a \Theta^{\be]}_b \\
t^{\al}_a & \equiv & \theta^\al_a - \Theta^\al_a;
\eeqa
the superfield properties of these are easily verified. The superfield $s^{\al\be}$ is a vector and we will, when appropriate, use the alternative notation $s^\mu$ for this, employing an $SO(5,1)$ vector index as in the bosonic case. Some important quantities will be stated twice, using both conventions.

We will also need the differential
\beq
ds^{\al\be} = dx^{\al\be} - i \Om^{ab} d\theta^{[\al}_a
  \Theta^{\be]}_b = e^{\al\be} - i \Om^{ab} (\theta-\Theta)^{[\al}_a
  d\theta^{\be]}_b,
\eeq
where the second equation follows from the definition of the
superspace differential $e^{\al\be}$. Similarly, we may differentiate
$s^{\al\be}$ with respect to the parameters defining the string
world-sheet, yielding
\beq
\tilde{d} s^{\al\be} = - \tilde{d} X^{\al\be} - i \Om^{ab}
\theta^{[\al}_a \tilde{d} \Theta^{\be]}_b = - E^{\al\be} - i
\Om^{ab} (\theta-\Theta)^{[\al}_a \tilde{d} \Theta^{\be]}_b,
\eeq
where $E^{\al\be}$ as before is the bosonic superspace differential expressed in terms of the embedding fields $X^{\al\be}$ and $\Theta^\al_a$.

Guided by \Eqnref{desi}, we introduce
\beqa
\eqnlab{Desi_1}
\De_{\sss \Si} & = & \frac{1}{4!} \int_\Si ds^\mu \we ds^\nu \we ds^\rho \we ds^\si \,\, \tilde{d}s^\tau \we \tilde{d}s^\veps \eps_{\mu\nu\rho\si\tau\veps} \de^{(6)}(s) = \\
\eqnlab{Desi_2}
& = & \frac{1}{4!} \int_{\Si} ds^{\de_1 \de_2} \we
ds^{\ga_1 \ga_2} \we ds^{\be_1 \be_2} \we ds^{\al_1 \al_2} \, \tilde{d}
s^{\si_1 \si_2} \we \tilde{d} s^{\tau_1 \tau_2} \times \nn \\
& & \qquad {} \times \eps_{\al_1 \al_2
  \be_1 \ga_1} \eps_{\ga_2 \de_1 \de_2 \si_1} \eps_{\si_2 \tau_1
  \tau_2 \be_2} \de^{(6)}(s),
\eeqa
where the Dirac delta function with a grassmannian argument containing both ``body'' and ``soul'' is defined in terms of its Taylor expansion. It is apparent that this candidate for $\De_{\sss \Si}$ reduces to the bosonic $\desi$ in \Eqnref{desi} if all fermions are put to zero.

The second requirement on this $\De_{\sss \Si}$ is that it should be closed, as is seen immediately from \Eqnref{calF}. The proof for this is similar to the bosonic case in \Eqnref{ddesi=0}, but slightly more complicated since $d\tilde{d} s^\mu = \tilde{d}ds^\mu \neq 0$. However, the changes in the proof are minor and therefore omitted in this text.

We also want to investigate how $\De_{\sss \Si}$ transforms under superconformal transformations. As in the bosonic case, we will present the transformations piece by piece. Using the explicit variations of $X^{\al\be}$ and $\Theta^\al_a$ found above, we find that
\beqa
\de_{\sss C} ( d s^{\al\be} ) & = & \de_{x,\theta} ds^{\al\be} - \tilde{\La}
d s^{\al\be} + 2 \dia{\tilde{\Om}}{\ga}{[\al} d s^{\be]\ga} + d \chi^{\al\be} \\
\de_{\sss C} ( \tilde{d} s^{\al\be} ) & = & \de_{x,\theta} \tilde{d}s^{\al\be} - \tilde{\La} \tilde{d} s^{\al\be} + 2 \dia{\tilde{\Om}}{\ga}{[\al}
  \tilde{d} s^{\be]\ga} + \tilde{d} \chi^{\al\be},
\eeqa
where we have used the presence of a Dirac delta function in the expression \eqnref{Desi_2} to put $s^{\al\be}=0$. Moreover,
$\de_{x,\theta}$ denotes a passive variation in superspace and
\beqa
\dia{\tilde{\Om}}{\ga}{\al} & \equiv & \dia{\Om}{\ga}{\al}(x,\theta) +
i \Om^{cd}\Xi_{c,\ga}(\theta)t_d^{\al} -
\frac{i}{4}\Om^{cd}\Xi_{c,\de}(\theta)t_d^{\de} \kde{\ga}{\al} \\
\tilde{\La} & \equiv & \La(x,\theta) + \frac{i}{2} \Om^{cd}
\Xi_{c,\ga}(\theta)t_d^\ga \\
\chi^{\al \be} & \equiv & -\Om^{ab}\Om^{cd} \Xi_{a,\ga}(\theta)
t_b^{[\al} t_c^{\be]} t_d^\ga - c_{\ga\de} \Om^{ab}\Om^{cd}
t_a^\ga t_b^{[\al} t_c^{\be]} t_d^\de.
\eeqa
Alternatively, we may write these variations as
\beqa
\de_{\sss C} ( d s^{\mu} ) & = & \de_{x,\theta} ds^{\mu} - \tilde{\La}
d s^{\mu} - \tilde{\Om}^{\mu}_{\ph{\mu}\nu} ds^\nu + d \chi^{\mu} \\
\de_{\sss C} ( \tilde{d} s^{\mu} ) & = &  \de_{x,\theta} \tilde{d}s^{\mu} - \tilde{\La} \tilde{d} s^{\mu} - \tilde{\Om}^{\mu}_{\ph{\mu}\nu} \tilde{d}s^\nu + \tilde{d} \chi^{\mu},
\eeqa
where we again have introduced a vector index instead of an antisymmetric pair of $SO(5,1)$ spinor indices. We have also introduced $\tilde{\Om}^{\mu}_{\ph{\mu}\nu}$, the definition of which is apparent from how $\dia{\om}{\al}{\be}$ was defined from $\om^{\mu\nu}$ in \Secref{ferm_conv}.

The Dirac delta function transforms according to
\beq
\de_{\sss C} (\de^{(6)}(s)) = \de_{x,\theta} \de^{(6)}(s) + 6 \tilde{\La} \de^{(6)}(s) + \chi \cdot \pa \de^{(6)}(s),
\eeq
where we as usual have used the properties of the delta function to
put $s^{\al\be}=0$. We have also employed the identity
\beq
s^{\al\ga} \pa_{\al\be} \de^{(6)}(s) = - \frac{3}{2}
\kde{\be}{\ga}\de^{(6)}(s),
\eeq
which is analogous to \Eqnref{delta_x} in the bosonic model.

Putting all this together, using the vector index notation, we find that the superconformal variation of $\De_{\sss \Si}$ is
\beqa
\de_{\sss C} \De_{\sss \Si} & = & \de_{x,\theta} \De_{\sss \Si} + \frac{1}{3!} \int_\Si ds^\mu \we ds^\nu \we ds^\rho \we d\chi^\si \,\, \tilde{d}s^\tau \we \tilde{d}s^\veps \eps_{\mu\nu\rho\si\tau\veps} \de^{(6)}(s) + {} \nn \\
& & {} + \frac{2}{4!} \int_\Si ds^\mu \we ds^\nu \we ds^\rho \we ds^\si \,\, \tilde{d}s^\tau \we \tilde{d}\chi^\veps \eps_{\mu\nu\rho\si\tau\veps} \de^{(6)}(s) + {} \nn \\
& & {} + \frac{1}{4!} \int_\Si ds^\mu \we ds^\nu \we ds^\rho \we ds^\si \,\, \tilde{d}s^\tau \we \tilde{d}s^\veps \eps_{\mu\nu\rho\si\tau\veps} \chi^\ka \pa_\ka  \de^{(6)}(s),
\eeqa
which may be rewritten as (our conventions for superderivatives may be found in~\cite{AFH:2003coupled})
\beqa
\de_{\sss C} \De_{\sss \Si} & = & \de_{x,\theta} \De_{\sss \Si} + d \Big[ \frac{1}{3!} \int_\Si ds^\mu \we ds^\nu \we ds^\rho \,\, \tilde{d}s^\tau \we \tilde{d}s^\veps \chi^\si \eps_{\mu\nu\rho\si\tau\veps} \de^{(6)}(s) \Big] - {} \nn \\
& & {} - \frac{2}{4!} \int_\Si \tilde{d} \left[ ds^\mu \we ds^\nu \we ds^\rho \we ds^\si \,\, \tilde{d}s^\tau \chi^\veps \eps_{\mu\nu\rho\si\tau\veps} \de^{(6)}(s) \right] + {} \nn \\
& & {} + \frac{7}{4!} \int_\Si ds^\mu \we ds^\nu \we ds^\rho \we ds^\si \,\, \tilde{d}s^\tau \we \tilde{d}s^\veps \chi^\ka \eps_{[\mu\nu\rho\si\tau\veps} \pa_{\ka]}  \de^{(6)}(s).
\eeqa
In this expression, the last line vanishes since it is an antisymmetrization over seven vector indices in six dimensions, while the second line is zero since it reduces to a boundary term. We are left with
\beq
\eqnlab{deDesi}
\de_{\sss C} \De_{\sss \Si} = \de_{x,\theta} \De_{\sss \Si} + d \La_{\sss \Si},
\eeq
where the super three-form $\La_{\sss \Si}$ is given by
\beqa
\La_{\sss \Si} & = & \frac{1}{3!} \int_\Si ds^\mu \we ds^\nu \we ds^\rho \,\, \tilde{d}s^\tau \we \tilde{d}s^\veps \chi^\si \eps_{\mu\nu\rho\si\tau\veps} \de^{(6)}(s) = \\
& = & \frac{1}{3!} \int_{\Si} ds^{\ga_1 \ga_2} \we
ds^{\be_1 \be_2} \we ds^{\al_1 \al_2} \chi^{\veps_1
  \veps_2} \tilde{d} s^{\si_1 \si_2} \we \tilde{d}
s^{\tau_1 \tau_2} \de^{(6)}(s) \times {} \nn \\
 & & \quad {} \times \Big( \eps_{\veps_1 \veps_2 \be_1 \ga_1} \eps_{\ga_2
  \al_1 \al_2 \si_1} \eps_{\si_2 \tau_1 \tau_2 \be_2} - \frac{1}{8}
\eps_{\veps_1 \veps_2 \ga_1 \ga_2} \eps_{\si_1 \si_2 \al_1 \al_2}
 \eps_{\tau_1 \tau_2 \be_1 \be_2} \Big),
\eeqa
but this quantity is obviously only well-defined modulo an exact three-form.

Looking back at \Eqnref{de=de} for the bosonic model, we see that the introduction of fermionic degrees of freedom has altered the relation by adding a $d$-exact term to the right-hand side. This means that our candidate \eqnref{Desi_1} for $\De_{\sss \Si}$ does not transform exactly as one would expect when comparing with the bosonic theory, \ie an active and a passive transformation do not yield the same result. We note that $\La_{\sss \Si}$ vanishes if all fermions are removed and that it is localized (by means of the Dirac delta function) to the world-sheet of the string.

This means that the simple generalization of the bosonic case that we have tried in this subsection did not work out properly. The most probable reason for this failure is that we are required to add a new ingredient, the matrix $\hat{\phi}^{ab}$, to $\De_{\sss \Si}$. This denotes the vacuum expectation value of the field $\phi^{ab}(x)$, normalized to unit modulus with respect to the scalar product in~\Eqnref{so5scalar}. In other words, $\hat{\phi}^{ab}$ is related to the moduli parameters of the theory, denoting the direction in which $R$-symmetry is broken by the presence of the tensile string. We have seen the appearance of this quantity previously~\cite{AFH:2003free} in the discussion concerning the $\kappa$-symmetry of the theory. We will not consider how to formulate a superspace generalization of the Poincar\'e dual including $\hat{\phi}^{ab}$ in this paper, but we hope to return to the matter in a future publication.

\vspace{7mm}
\noindent
\textbf{Acknowledgments:}
I would like to thank Erik Flink and M{\aa}ns Henningson for
stimulating and encouraging discussions.

\clearpage
\bibliographystyle{utphysmod3b}
\addcontentsline{toc}{section}{References}
\bibliography{biblio}

\end{document}